\documentclass[twocolumn, showpacs, preprintnumbers, nofootinbib, aps, prd, superscriptaddress, 10pt, showkeys, notitlepage]{revtex4-1}
\usepackage[utf8]{inputenc}
\usepackage{graphicx,amssymb,amsmath,amsthm,amsfonts,epsfig,times,natbib}

\usepackage[linktocpage,breaklinks]{hyperref}
\usepackage[usenames,dvipsnames]{color}
\usepackage{epstopdf}
\usepackage{aas_macros,amsmath,amssymb}
\usepackage{tensor}
\usepackage{mathtools}
\usepackage{amsbsy}
\usepackage{bm}

\definecolor{oxfordblue}{rgb}{0.0, 0.13, 0.48}
\definecolor{burgundy}{rgb}{0.5, 0.0, 0.13}
\definecolor{crimsonglory}{rgb}{0.75, 0.0, 0.2}
\definecolor{darkolivegreen}{rgb}{0.33, 0.42, 0.18}
\definecolor{darkblue}{rgb}{0,0,0.5}
\definecolor{richcarmine}{rgb}{0.84, 0.0, 0.25}
\definecolor{darkblue}{rgb}{0,0,0.5}
\definecolor{venetianred}{rgb}{0.78, 0.03, 0.08}
\definecolor{skobeloff}{rgb}{0.0, 0.48, 0.45}
\hypersetup{colorlinks=true, citecolor=darkblue, linkcolor=darkblue,
urlcolor = darkblue}

\newcommand{\be}{\begin{equation}}
\newcommand{\ee}{\end{equation}}
\newcommand{\bear}{\begin{eqnarray}}
\newcommand{\eear}{\end{eqnarray}}

\newcommand{\nn}{\nonumber}

\newcommand{\p}{\partial}

\newcommand{\baa}{\begin{align}}
\newcommand{\eaa}{\end{align}}

\newcommand{\bea}{\begin{eqnarray}}
\newcommand{\eea}{\end{eqnarray}}

\begin{document}
\title{Multipole moments and universal relations for scalarized neutron stars}

\begin{abstract}
In recent years there has been a surge of interest in what has come to be known as the ``universal relations'' between various global properties of neutron stars. These universal relations are equation of state independent relations between quantities such as the moment of inertia $I$, the tidal deformability or Love number $\lambda$, and the quadrupole $Q$ (I-Love-Q relations), or the relativistic multipole moments (3-hair relations).  While I-Love-Q relations  have been studied extensively in both general relativity and various alternatives, 3-hair relations have been studied only in general relativity. Recent progress on the definition of the multipole moments of a compact object in the case of scalar-tensor theories allows for the study of 3-hair relations in modified theories of gravity. Specifically, the aim of this work is to study them for scalarized stars in scalar-tensor theories with a massless scalar field that admit spontaneous scalarization. We find that the 3-hair relations between the mass and angular momentum moments that hold in general relativity  hold for scalarized stars as well. The scalar moments also exhibit a universal behaviour, which is equation of state independent within one specific theory, but differs between different theories. Combining astrophysical observations one can in principle measure the different properties of scalarized neutron star and tell different theories apart. 
\end{abstract}

\author{George Pappas}
\email{georgios.pappas@roma1.infn.it}
\affiliation{Dipartimento di Fisica, ``Sapienza'' Universit\'a di Roma \& Sezione INFN Roma1, Piazzale Aldo Moro 5, 00185, Roma, Italy}
\affiliation{School of Mathematical Sciences,  University of Nottingham, University Park, Nottingham NG7 2RD, UK}

\author{Daniela D. Doneva}
\email{daniela.doneva@uni-tuebingen.de}
\affiliation{Theoretical Astrophysics, Eberhard Karls University of T\"ubingen,
T\"ubingen, D-72076, Germany}
\affiliation{INRNE - Bulgarian Academy of Sciences, 1784 Sofia, Bulgaria}

\author{Thomas P. Sotiriou}
\email{Thomas.Sotiriou@nottingham.ac.uk}
\affiliation{School of Mathematical Sciences,  University of Nottingham,
University Park, Nottingham NG7 2RD, UK}
\affiliation{School of Physics and Astronomy, University of Nottingham,
University Park, Nottingham NG7 2RD, UK}

\author{Stoytcho S. Yazadjiev}
\email{yazad@phys.uni-sofia.bg}
\affiliation{Department of Theoretical Physics, Faculty of Physics, Sofia University, Sofia 1164, Bulgaria}
\affiliation{Theoretical Astrophysics, Eberhard Karls University of T\"ubingen,
T\"ubingen, D-72076, Germany}
\affiliation{Institute of Mathematics and Informatics, 	Bulgarian Academy of Sciences, 	Acad. G. Bonchev St. 8, Sofia 1113, Bulgaria}

\author{Kostas D. Kokkotas}
\email{kostas.kokkotas@uni-tuebingen.de}
\affiliation{Theoretical Astrophysics, Eberhard Karls University of T\"ubingen,
T\"ubingen, D-72076, Germany}

\date{{\today}}


\maketitle

 \tableofcontents
 
\section{Introduction}
\label{sec:intro}

Neutron stars have been studied extensively in scalar-tensor theories (STT) of gravity. In contrast to black holes that do not possess scalar hair in these theories \cite{Hawking:1972qk,Bekenstein:1995un,Sotiriou:2011dz}, for  neutron stars the  matter acts as a source of the scalar field and supports nontrivial scalar configurations. The simplest class of STT, Brans-Dicke theory, always leads to the development of a nontrivial scalar field for all matter configurations and the differences with  general relativity (GR) can be considerable. This is actually a disadvantage, as  Brans-Dicke theory deviates from Einstein's theory of gravity in the weak field regime where GR is tested with very high precision. Hence, weak-field test can already place tight constraints that leave little room for strong-field deviations.  This argument can be circumvented in a specific class of theories  that are perturbatively equivalent to GR in the weak field regime but exhibit significant deviations in the strong field regime  \cite{Damour:1993hw}.  In particular, neutron stars in such theories can exhibit spontaneous scalarization.\footnote{ As a matter of fact very similar phenomenon of scalarization is observed also for black holes in scalar-tensor theories in the presence of nonlinear fields \cite{Stefanov2008,Doneva2010a,Cardoso2013,Kleihaus2015}.} The essence of  spontaneous scalarization is that, once a star exceeds a compactness threshold, it is energetically more favorable to develop a nontrivial scalar configuration \cite{Damour:1993hw,Harada:1998ge}. Scalarized neutron stars have been examined further  \cite{Sotani:2017pfj,Motahar:2017blm} including slow  \cite{Damour:1996ke,Sotani:2012eb,Pani:2014jra,Silva:2014fca} and rapid \cite{Doneva:2013qva,Doneva:2014uma} rotation.

As in GR, the properties of neutron stars in STT depend on the specific equation of state (EOS) that one selects to describe their interior.  The various uncertainties in the microphysics  result in a proliferation of EOSs. The properties of the star and the relevant observables depend both on the choice of the EOS and on the theory of gravity and these dependencies appear to be degenerate. That is, using a different theory of gravity, such as STT, is hard to distinguish from changing the EOS.   A way around this problem has already been proposed and continues to be developed in the form of universal, or EOS independent, relations between various quantities that characterise the structure of neutron stars (for some recent reviews see \cite{YagiYunes2016arXiv,Doneva2017review}). Such universal relations have  been studied in STT as well, though not as widely as in GR. A particular class of the universal relations that has been studied in GR but not in STT are the 3-hair relations between the multipole moments of a neutron star (for the results in GR see \cite{Pappas2014PhRvL,Stein2014ApJ,Yagi2014PhRvD}). The goal of this work is to extend the GR results on 3-hair relations to the case of scalarized stars and to explore the possibility of having a universal description of the moments of the additional degree of freedom, i.e., the scalar field. 
We will also explore how the scalar moment are related to observables and whether we can use such observables to put constraints on the parameters of the STT, as well as the degree of scalarization of a neutron star.

Since the class of STT discussed above cannot be constrained by the weak field experiments, one has to use observations involving strong field effects, such as the gravitational wave emission of neutron stars located in close binary systems leading to shrinking of their orbits. These observations pose strong constraints on the theory \cite{Damour:1996ke} and the latest results lead to tight bounds on the free coupling parameters \cite{2010Natur.467.1081D,Antoniadis:2013pzd}. Thus, the nonrotating and slowly rotating scalarized neutron stars do not differ significantly from the GR case and it would be very difficult to probe the presence of a scalar field. Only the rapidly rotating case can lead to larger deviations from the non-scalarized solutions \cite{Doneva:2013qva}. This motivates an investigation of the universal relations for rapidly rotating stars, which is the focus of this paper. 

We should note that there is another way to evade the strong constraints coming from the binary pulsar observations, namely the inclusion of nonzero scalar field mass. This would lead to a finite range for the scalar field of the order of its Compton wavelength and can reconcile the theory with  observations for a much larger range of parameters. Nonrotating neutron stars in massive STT were examined for the first time in \cite{Popchev2015,Ramazanoglu:2016kul}  and the results were extended in \cite{Yazadjiev:2016pcb} and \cite{Doneva:2016xmf} for slow and rapid rotation respectively. The studies indeed showed that the neutron stars can differ dramatically from the pure general relativistic case. Defining the multipole moments in these theories is more complicated though and we will leave it for future studies. 

The rest of this work is organised in the following way, Section \ref{sec:setup} gives a brief description of the formalism for construction neutrons stars in STT, while Section \ref{sec:moments} gives a brief description of the calculation of the moments. Section \ref{sec:universal} presents the results on the various universal relations between the multipole moments, and Section \ref{sec:observables} discusses how the various relations could be used to extract information about the moments and the particular STT from observations. Finally, we end with our conclusions.

\section{Stars in scalar-tensor theory}
\label{sec:setup}

The general form of the Einstein frame action in STTs with a massless scalar field is \cite{Fujii2003,Will2006,Damour1992}
\begin{eqnarray}
S&=& {1\over 16\pi G_{*}}\int d^4x \sqrt{-g} \left(R -
2g^{\mu\nu}\partial_{\mu}\varphi \partial_{\nu}\varphi \right)\nonumber\\
   &&+ S_{m}[\Psi_{m}; {\cal A}^{2}(\varphi)g_{\mu\nu}],
\end{eqnarray}
where $G_{*}$ is the bare gravitational constant, $R$ is the Ricci scalar curvature with respect to the Einstein frame metric $g_{\mu\nu}$, the matter fields are collectively denoted by $\Psi_{m}$ and their action is $S_{m}$. In the Einstein frame the scalar field $\varphi$ is directly coupled to the matter via the function  ${\cal A}(\varphi)$. This function plays the role of a conformal factor that relates the Einstein frame metric $g_{\mu\nu}$ to the  Jordan frame metric $\tilde{g}_{\mu\nu}={\cal A}^{2}(\varphi)g_{\mu\nu}$. By definition, the matter fields couple minimally to the Jordan frame metric and this guarantees that the weak equivalence principle is satisfied. We have chosen to work in the Einstein frame, as in this frame the field equations have the same structure as in GR and this simplifies calculations. Moreover, the multipole moments presented below have been previously defined and  calculated in the Einstein frame \cite{Pappas:2015moments}. We stress that any physical quantities in the Jordan frame can be expressed in terms of these moments \cite{PappasMNRAS2015ST}. 

In what follows we use geometrical units $c=G_{* }=1$ and the dimensional quantities are given in $\rm km$. We will focus on stellar configurations that are stationary and axisymmetric and we will describe the matter in the Einstein frame as a perfect fluid with pressure $p$ and energy density $\varepsilon$. The spacetime metric can then 
 be written in the following general form 
\begin{eqnarray}
ds^2 = -e^{\gamma+\sigma} dt^2 &+& e^{\gamma-\sigma} r^2
\sin^2\theta (d\phi - \omega dt)^2 \nonumber\\
   &+& e^{2\alpha}(dr^2 + r^2
d\theta^2)\,. \label{eq:metric_RNS}
\end{eqnarray}
All metric functions $\gamma$, $\sigma$, $\omega$ and $\alpha$, as well  $\varphi$, $p$ and $\varepsilon$, will depend only on the coordinates $r$ and $\theta$. For numerical calculations it is more convenient to use the angular coordinate $\mu=\cos\theta$ instead  of $\theta$. Using our ans\"atze, the field equations that one obtains from varying the action with respect to the metric take the form
\begin{widetext}
\begin{eqnarray}
\left(\Delta + \frac{1}{r} \partial_{r}  -
\frac{\mu}{r^2}\partial_{\mu}\right)\left(\gamma
e^{\gamma/2}\right)&=&e^{\gamma/2}\left\{16\pi p e^{2\alpha}  +\frac{\gamma}{2}\left[16\pi p  e^{2\alpha} -
\frac{1}{2}(\partial_{r}\gamma)^2 - \frac{1}{2} \frac{1-\mu^2}{r^2}
(\partial_{\mu}\gamma)^2 \right]\right\}, \label{eq:DiffEq_gamma}
\end{eqnarray}

\begin{eqnarray}
\Delta(\sigma e^{\gamma/2}) &=& e^{\gamma/2}\left\{8\pi (\varepsilon
+ p)e^{2\alpha} \frac{1+ \upsilon^2}{1 - \upsilon^2}   + r^2
(1-\mu^2)e^{-2\sigma}\left[(\partial_{r}\omega)^2  +
\frac{1-\mu^2}{r^2}(\partial_{\mu}\omega)^2\right] +
\frac{1}{r}\partial_{r}\gamma - \frac{\mu}{r^2}\partial_{\mu}\gamma
\right. \nonumber  \\ && \nonumber \\
&&\left.  + \frac{\sigma}{2}\left[16\pi p  e^{2\alpha} - \frac{1}{r}\partial_{r}\gamma + \frac{\mu}{r^2}\partial_{\mu}\gamma -
\frac{1}{2}(\partial_{r}\gamma)^2  - \frac{1}{2}\frac{1-\mu^2}{r^2} (\partial_{\mu}\gamma)^2 \right] \right\}, \label{eq:DiffEq_sigma}
\end{eqnarray}

\begin{eqnarray}
\left(\Delta  +  \frac{2}{r} \partial_{r}  -
\frac{2\mu}{r^2}\partial_{\mu}\right)\left(\omega e^{\gamma/2
	-\sigma}\right) &=& e^{\gamma/2 - \sigma} \left\{- 16\pi
\frac{(\varepsilon + p)(\Omega - \omega)}{1- \upsilon^2}e^{2\alpha}
+ \right. \nonumber \\
&&\omega\left[-\frac{1}{r}\partial_{r} (\frac{1}{2}\gamma + 2\sigma)
+ \frac{\mu}{r^2}\partial_{\mu}(\frac{1}{2}\gamma + 2\sigma) -
\frac{1}{4}(\partial_{r}\gamma)^2 -  \frac{1}{4}\frac{1-\mu^2}{r^2}
(\partial_{\mu}\gamma)^2  + \right. \nonumber \\
&& + (\partial_{r}\sigma)^2 + \frac{1-\mu^2}{r^2}
(\partial_{\mu}\sigma)^2     - r^2(1-\mu^2)e^{-2\sigma}
\left((\partial_{r}\omega)^2 + \frac{1-\mu^2}{r^2}
(\partial_{\mu}\omega)^2\right)  \nonumber \\
&& \left.  \left. - 8\pi \frac{\varepsilon (1+ \upsilon^2) +
	2p\upsilon^2}{1- \upsilon^2} e^{2\alpha} \right] \right\},\label{eq:DiffEq_omega}
\end{eqnarray}

\begin{eqnarray}
&&\partial_\mu \alpha = -\frac{\partial_\mu \gamma + \partial_\mu \sigma}{2} - \left\{(1-\mu^2)(1+r\partial_r\gamma)^2 + [-\mu + (1-\mu^2)\partial_\mu \gamma]^2 \right\}^{-1} \times  \label{eq:DiffEq_alpha}\\ \notag \\
&&\left\{\frac{1}{2}\left[r\partial_r(r\partial_r\gamma) + r^2 (\partial_r \gamma)^2 - (1-\mu^2)(\partial_\mu \gamma)^2 - \partial_\mu[(1-\mu^2)\partial_\mu \gamma] + \mu\partial_\mu \gamma\right] \times [-\mu + (1-\mu^2)\partial_\mu \gamma] + \right. \notag \\ \notag \\
&& +\frac{1}{4}[-\mu + (1-\mu^2)\partial_\mu \gamma] \times \left[r^2(\partial_r \gamma + \partial_r \sigma)^2 - (1-\mu^2)(\partial_\mu \gamma + \partial_\mu \sigma)^2 + 4r^2(\partial \varphi)^2 - 4(1-\mu^2)(\partial_\mu \varphi)^2\right] + \notag \\ \notag \\
&& + \mu r \partial_r\gamma [1+r\partial_r\gamma] - (1-\mu^2)r(1+r\partial_r\gamma)\left[\partial_\mu\partial_r \gamma + \partial_\mu\gamma\partial_r\gamma + \frac{1}{2}(\partial_\mu \gamma + \partial_\mu \sigma)(\partial_r \gamma + \partial_r \sigma) + 2 \partial_\mu\varphi\partial_r \varphi \right] +\notag \\ \notag \\
&&+\frac{1}{4}(1-\mu^2)e^{-2\sigma}\left[-[-\mu+(1-\mu^2)\partial_\mu
\gamma][r^4(\partial_r \omega)^2 - r^2(1-\mu^2)(\partial_\mu
\omega)^2] + \right. \notag \\ \notag \\
&&\left.\left.+2(1-\mu^2)r^3\partial_\mu\omega\partial_r\omega(1+r\partial_r\gamma)\right]\right\},
\notag
\end{eqnarray}
\end{widetext}
where the differential operator $\Delta$ is defined as
\begin{eqnarray}
\Delta =  \partial^2_{r} + \frac{1}{r}\partial_{r} +
\frac{1-\mu^2}{r^2}\partial^2_{\mu} - \frac{2\mu}{r^2}
\partial_{\mu}.
\end{eqnarray}
The last field equation \eqref{eq:DiffEq_alpha} for the metric function $\alpha$  is of first order compared to the second order field equations for the rest of the metric functions.
The field equation for the scalar field is 
\begin{eqnarray}
\Delta \varphi&=& - \partial_{r}\gamma\partial_{r}\varphi -
\frac{1-\mu^2}{r^2} \partial_{\mu}\gamma\partial_{\mu}\varphi \nn \\
&&+ 4\pi k(\varphi)(\varepsilon - 3p) e^{2\alpha}.
\label{eq:DiffEq_phi}
\end{eqnarray}

The above system of  equations has to be supplemented with equations that describe the dynamics of the fluid, namely the equation for hydrostatic equilibrium and the equation of state (EOS) for nuclear matter. The latter is a relation between pressure and energy density and we impose it in the Jordan frame, since the matter couples minimally to the Jordan frame metric. This minimal coupling also implies that the fluid will satisfy the usual conservation laws in terms of the Jordan frame variable, $\tilde{p}$ and $\tilde{\varepsilon}$. The equations above  have been given in the Einstein frame and $p$ and $\varepsilon$ are related to $\tilde{p}$ and $\tilde{\varepsilon}$ as follows
\begin{eqnarray}
\label{EOSconf}
\varepsilon &=&{\cal A}^4(\varphi){\tilde\varepsilon}, \nonumber \\
p&=&{\cal A}^4(\varphi){\tilde p}.
\end{eqnarray}
One can use these relations to express the EOS and the equation for hydrostatic equilibrium in terms of $p$ and $\varepsilon$ in order to work exclusively with Einstein frame variables. We find it more convenient to work directly with $\tilde{p}$ and $\tilde{\varepsilon}$. Hence, in the numerical implementation we use eqs.~\eqref{EOSconf} to express $p$ and $\varepsilon$ in terms  $\tilde{p}$ and $\tilde{\varepsilon}$ in all the equations above and we express the equation for hydrostatic equilibrium in the form 
\begin{eqnarray}
\frac{\partial_i{\tilde p}}{{\tilde \varepsilon} + \tilde{p}} -
\left[\partial_i(\ln \, u^t) - u^t u_\phi \partial_i \Omega -
k(\varphi) \partial_i \varphi\right]=0 \label{eq:Hydrostatic_Equil},
\end{eqnarray}
where we have introduced the coupling function $k(\varphi)= {d\ln({\cal  A}(\varphi))}/ {d\varphi}$. The Einstein frame four velocity $u^\mu$ is defined as
\begin{equation}
u^\mu = \frac{e^{-(\sigma + \gamma)/2}}{\sqrt{1-v^2}}
[1,0,0,\Omega],
\end{equation}
where  the proper velocity is $v = (\Omega - \omega) r \sin \theta e^{-\sigma}$ and $\Omega$ is the fluid angular velocity $\Omega={u^{\phi}}/{u^{t}}$.

What is left to be fixed then is the particular form of the Einstein frame coupling function. We will work with the standard choice $k(\varphi)=\beta \varphi$ where $\beta$ is a constant. One of the most important properties of this class of scalar-tensor theories is that it is perturbatively equivalent to GR in the weak field regime, while in the strong field regime nonlinear effects lead to  non-uniqueness of  solutions and spontaneous scalarization \cite{Damour:1993hw}. In the calculations below we will allow also for nonzero background value of the scalar field $\varphi_{\infty} $ in some cases.

We solve the field equations using a modification of the {\tt RNS} code (see \cite{Stergioulas95} for the original GR version of the {\tt RNS} code while the STT extension can be found in \cite{Doneva:2013qva}). This code is based on the KEH method \cite{Komatsu:1989zz} with certain modifications introduced in \cite{Cook1992}. A key property of this method is that the field equations are presented in an integral form. This turns out to be  very useful for the calculation of the multipole moments,  as explained in Appendix \ref{sec:app2}.

\section{Mass, angular momentum and scalar field moments in scalar-tensor theory}
\label{sec:moments}

Here we give a brief description of the framework and the general results for the moments in the Einstein frame \cite{Pappas:2015moments} for a STT with a massless scalar field. More details on the particular calculation of the moments employed in the {\tt RNS} code can be found in Appendix \ref{sec:app2}, while a general review of the calculation in GR can be found in \cite{Doneva2017review}.

When discussing the multipole moments it is more convenient to use the following form of the metric that is written again in quasi-isotropic coordinates similar to the metric used by the {\tt RNS} code \eqref{eq:metric_RNS}, but with the new functions $B=e^{\gamma}$ and $\nu=(\gamma+\sigma)/2$, i.e.,
\bear ds^2=-e^{2\nu}dt^2+r^2\sin^2\theta B^2 e^{-2\nu}(d\varphi-\omega dt)^2\nn\\
+e^{2\alpha}(dr^2+r^2d\theta^2). \label{eq:metric_MultipoleMoments}
\eear
The field equations for this metric are directly related to the ones given in the previous section, i.e., eqs. \eqref{eq:DiffEq_gamma}--\eqref{eq:DiffEq_alpha}.
Note that 
 the Einstein frame field equations \eqref{eq:DiffEq_gamma}--\eqref{eq:DiffEq_omega} are identical to their GR counterparts\footnote{This is true only if there is no potential for the scalar field.} (given in \cite{Cook1992}), while eq. \eqref{eq:DiffEq_alpha} and the equation for hydrostationary equilibrium \eqref{eq:Hydrostatic_Equil} have some additional contributions involving derivatives of the scalar field $\partial_i\varphi$. Therefore, as discussed in more detail in \cite{Doneva:2013qva,Pappas:2015moments}, in the case of a massless scalar field the multipole moments can be calculated in the same way as in GR with scalar field corrections entering through the Ricci tensor and the equation for $\alpha$, i.e., eq. \eqref{eq:DiffEq_alpha}. 
Similarly, 
the vacuum field equations for the metric functions $\nu$, $B$ and $\omega$, which are used to define the moments, are the same as in GR and can be found  in \cite{BI1976ApJ}. One can easily show that the asymptotic expansion of the metric functions and the scalar field admits the following ansatz in terms of the Legendre Polynomials $P_l(\mu)$, their derivatives $\frac{d P_l(\mu)}{d\mu}$, and the Gegenbauer polynomials $T_l^{1/2}(\mu)$,\footnote{We draw the reader's attention to the definition for the Gegenbauer polynomials given in \cite{BI1976ApJ}, which might be different in other sources in the literature.}  
\bea 
\nu&=&\sum_{l=0}^{\infty}\bar{\nu}_{2l}(r)P_{2l}(\mu),\\
\omega&=&\sum_{l=1}^{\infty}\bar{\omega}_{2l-1}(r)\frac{dP_{2l-1}(\mu)}{d\mu},\\
B&=&1+\sum_{l=0}^{\infty}\bar{B}_{2l}(r)T_{2l}^{1/2}(\mu),\\
\varphi & =& \sum_{n=0}^{\infty} \bar{\Phi}_{2n}(r)P_{2n}(\mu). 
\eea
where 
the coefficients in these expansions are of the form
\bea 
\bar{\nu}_{2l}(r)&=&\sum_{k=0}^{\infty}\frac{\nu_{2l,k}}{r^{2l+1+k}} \label{eq:Expansion_nu1}\\
\bar{\omega}_{2l-1}(r)&=&\sum_{k=0}^{\infty}\frac{\omega_{2l-1,k}}{r^{2l+1+k}}, \label{eq:Expansion_omega2}\\
\bar{B}_{2l}(r)&=&\frac{B_{2l}}{r^{2l+2}}, \label{eq:Expansion_B3} \\
\bar{\Phi}_{2n}(r) &=& \frac{\Phi_{2n}}{r^{2n+1}}. \label{eq:Expansion_phi4}
\eea

The calculated multipole moments of the spacetime are combinations of the expansion coefficient in \eqref{eq:Expansion_nu1}--\eqref{eq:Expansion_phi4} (see discussion for the GR case in \cite{Doneva2017review}) and below we give explicitly the first few multipole moments using the formalism developed in \cite{Pappas:2015moments}. We should note that even though the calculation of the metric coefficients \eqref{eq:Expansion_nu1}--\eqref{eq:Expansion_B3} is the same as in GR, the coefficients in the scalar field expansion enter explicitly in the multipole moments given below.

Mass (monopole):

\be M\equiv M_0=-\nu_{0,0} \ee

Scalar monopole: 

\be  W_0=\Phi_0 \ee

Angular momentum (dipole):

\be  J\equiv S_1=\frac{\omega_{1,0}}{2} \ee

Mass quadrupole:

\be  M_2=\frac{4}{3} B_0 \nu_{0,0}+\frac{1}{3} \Phi_0^2 \nu_{0,0}+\frac{\nu_{0,0}^3}{3}-\nu_{2,0} \ee

Scalar quadrupole:

\be W_2=-\frac{1}{3} \Phi _0 \nu _{0,0}^2-B_0 \Phi _0-\frac{\Phi _0^3}{3}+\Phi _2  \ee

Spin octupole:

\be  S_3=-\frac{6}{5} B_0 \omega_{1,0}-\frac{3}{10} \nu_{0,0}^2 \omega_{1,0}-\frac{3}{10} \Phi_0^2 \omega_{1,0} + \frac{3 \omega_{3,0}}{2} \ee

Mass hexadecapole:

\bear  M_4&=&-\frac{10}{7} B_0 \Phi _0^2 \nu _{0,0}-\frac{32}{21} B_0 \nu _{0,0}^3-\frac{16}{5} B_0^2 \nu _{0,0}\nonumber\\
   &&+\frac{64}{35} B_2 \nu
   _{0,0}+\frac{24}{7} B_0 \nu _{2,0}-\frac{38}{105} \Phi _0^2 \nu _{0,0}^3\nonumber\\
   &&-\frac{19}{105} \Phi _0^4 \nu _{0,0}+\frac{2}{7} \Phi _0 \Phi _2
   \nu _{0,0}+\frac{6}{7} \Phi _0^2 \nu _{2,0}\nonumber\\
   &&+\frac{3}{70} \nu _{0,0} \omega _{1,0}^2-\frac{19}{105} \nu _{0,0}^5+\frac{8}{7} \nu _{2,0} \nu
   _{0,0}^2-\nu _{4,0}
\eear

Scalar hexadecapole:

\bear  W_4&=&\frac{26}{21} B_0 \Phi _0 \nu _{0,0}^2+\frac{38}{105} \Phi _0^3 \nu _{0,0}^2+\frac{19}{105} \Phi _0 \nu _{0,0}^4\nonumber\\
   &&-\frac{2}{7} \Phi _0 \nu
   _{0,0} \nu _{2,0}-\frac{6}{7} \Phi _2 \nu _{0,0}^2-\frac{3}{70} \Phi _0 \omega _{1,0}^2\nonumber\\
   &&+\frac{8}{7} B_0 \Phi _0^3+2 B_0^2 \Phi _0-B_2 \Phi
   _0-3 B_0 \Phi _2\nonumber\\
   &&+\frac{19 \Phi _0^5}{105}-\frac{8}{7} \Phi _2 \Phi _0^2+\Phi _4 \eear

Spin $2^5$-pole: 

\bear S_5&=&\frac{104}{63} B_0 \nu_{0,0}^2 \omega_{1,0}+\frac{11}{7} B_0 \Phi_0^2 \omega_{1,0}+\frac{24}{7} B_0^2 \omega_{1,0}\nonumber\\
   &&-\frac{32}{21} B_2 \omega_{1,0}-\frac{20}{3} B_0 \omega_{3,0}+\frac{25}{63} \Phi_0^2 \nu_{0,0}^2 \omega_{1,0}\nonumber\\
   &&+\frac{25}{126} \nu_{0,0}^4 \omega_{1,0}-\frac{5}{21} \nu_{0,0} \nu_{2,0} \omega_{1,0}-\frac{5}{3} \nu_{0,0}^2 \omega_{3,0}\nonumber\\
   &&+\frac{25}{126} \Phi_0^4 \omega_{1,0}-\frac{5}{3} \Phi_0^2 \omega_{3,0}-\frac{5}{21} \Phi_2 \Phi_0 \omega_{1,0}\nonumber\\
   &&-\frac{\omega_{1,0}^3}{28}+\frac{5 \omega_{5,0}}{2} \eear

These are all the non-zero multipole moments up to $S_5$ for a stationary and axisymmetric spacetime with equatorial symmetry and in the presence of a scalar field with the same symmetries. 

As emphasized earlier already, these moments are the Einstein frame moments. Defining the moments in the Einstein frame is straightforward, while attempting to do so in the Jordan frame appears to be significantly harder.
The Jordan frame is related to the Einstein frame through a conformal transformation that depends on the scalar field. In terms of the multipole moments, a conformal transformation of the metric would generally result in a mixing of the moments, with the new moments being combinations of the old ones. This is clearly not an essential redefinition of the multipole moments. In our specific case we would additionally have the mixing of mass and angular momentum moments with scalar field moments, due to the conformal factor being a function of $\varphi$. 
Any physical quantity that one would wish to express in term of some Jordan frame moments (assuming that they can be rigorously defined), can be always reexpressed in terms of the Einstein frame moments, using the relations between Einstein and Jordan frame variables. 
 A further advantage of the Einstein frame moments is the following. In the context of STT the functional form of the conformal factor ${\cal A}^{2}(\varphi)$ is specific to a theory or a class of theories and can be parameterised in terms of appropriate parameters or coupling coefficients of the theory. In the   selected formulation, these coupling coefficients of a specific theory appear as the coefficients that mix the Einstein frame moments, instead of being hidden in some Jordan frame moments. This gives a more transparent handle on a specific STT (see for example \cite{PappasMNRAS2015ST}). 
 Therefore, while the choice made here does not lose in generality, it can be further argued to be multiply advantageous.

As a last note, we mention that below we will use the reduced moments defined as, 
\be \bar{M}_{2n}=(i)^{2n}\frac{M_{2n}}{j^{2n} M^{2n+1}},\ee 
and 
\be \bar{S}_{2n+1}=(i)^{2n}\frac{S_{2n+1}}{j^{2n+1} M^{2n+2}},\ee 
where $n\geq 0$, $i$ is the imaginary unit, and $j\equiv J/M^2$. %
A similar normalisation will be used for the scalar moments, but this will be further explained in the following section. 

\section{EoS independent behaviour of scalarized stars.}
\label{sec:universal}

To explore the existence of universal relations between the various moments of scalarized stars, similar to the 3-hair relations between the moments in GR, we have constructed sequences of scalarized models using various EOSs. For these stars we have calculated the mass and angular momentum moments up to $M_4$, as well as the scalar moments up to $W_4$, following the procedure outlined in the previous section and the expressions given there. 
While the mass and angular momentum moments can be directly compared to their GR counterparts, the scalar moments don't have a GR counterpart and are in this sense novel features. 

We use several equations of state in order to cover a wide range of stiffness. These are the APR4 \cite{Akmal:1998cf}, SLy4 \cite{Douchin:2001sv}, A \cite{Arnett:1976dh}, FPS \cite{Friedman:1981qw} and the zero temperature limit of the Shen EOS \cite{Shen:1998gq,Shen:1998by}.  APR4 and Sly4 are modern realistic EOS that are in agreement with the observations. 
EOS A and FPS are too soft and already excluded by  observations, as they do not reach two solar masses \cite{Demorest2010,Antoniadis:2013pzd}. The Shen EOS does reach the two solar mass barrier, but it is stiffer and leads to  somewhat larger radii, so it is disfavoured by observations \cite{Lattimer:2013hma,Ozel:2016oaf,Abbott:2018exr}. We have included softer and stiffer EOS even though they are ruled out or disfavoured, as our main goal is to demonstrate the universality of the relations given below. Hence, it is instructive to use a broader set of EOS in order to verify that this universality is not simply a residual effect from considering EOSs with very similar properties.

The scalarized models have been constructed assuming values of $\beta$ in the range between $-4$ and $-10$ covering a big part of the parameter space. We should note that the current observational limit is $\beta>-4.5$ \cite{Demorest2010,Antoniadis:2013pzd} for theories with a massless scalar field. Nevertheless, we have again decided to  include larger values of $|\beta|$, to demonstrate that the universality persists for significantly scalarized stars and it is not an artefact of very weak scalarization. It is worht mentioning that considering values of $\beta$ lower than $-4.5$ is justified if one includes a mass for the scalar field. 
In that case, the scalar field is confined within its Compton wavelength and therefore, for large enough scalar field masses, the emission of scalar gravitational radiation is suppressed and  binary pulsar observations cannot set as tight constraints on the parameter $\beta$ \cite{Popchev2015,Ramazanoglu:2016kul,Yazadjiev:2016pcb} as in the massless case. 
One should note however that defining the multipole moments in the case of massive scalar field is much more involved because of the finite range of the scalar field and its exponential decay at infinity. This remains an open problem which we plan on addressing in future work. One more issue we should address at this point is that of the asymptotic value of the scalar field, which in the class of models that we are investigating, is constrained to be almost zero by observations. Nevertheless, we have also calculated models with a non-zero asymptotic value of the scalar field $\varphi_0$  in order to have a more complete investigation of scalarized stars. For these latter models we have used a somewhat larger value of $\varphi_0$ (i.e., $\varphi_0 = 0.03$), similar to previous studies \cite{Sotani:2004rq}, in order to have a better assessment of how that would affect the behaviour of the universal relations. 
  
For the particular choice of the coupling function the field equations are invariant under the transformation $\varphi \rightarrow -\varphi$. Thus the neutron star solutions with opposite signs of the scalar field are otherwise indistinguishable (e.g. the metric functions describing the two solutions are the same). Therefore, in the presented results we have chosen arbitrarily one particular sign of $\varphi$ and normalised the scalar field multipole moments accordingly. In any case, solutions with the opposite sign for the scalar field also exist and would simply result to scalar moments with an opposite sign.

\begin{figure}[h]
\includegraphics[width=0.48\textwidth]{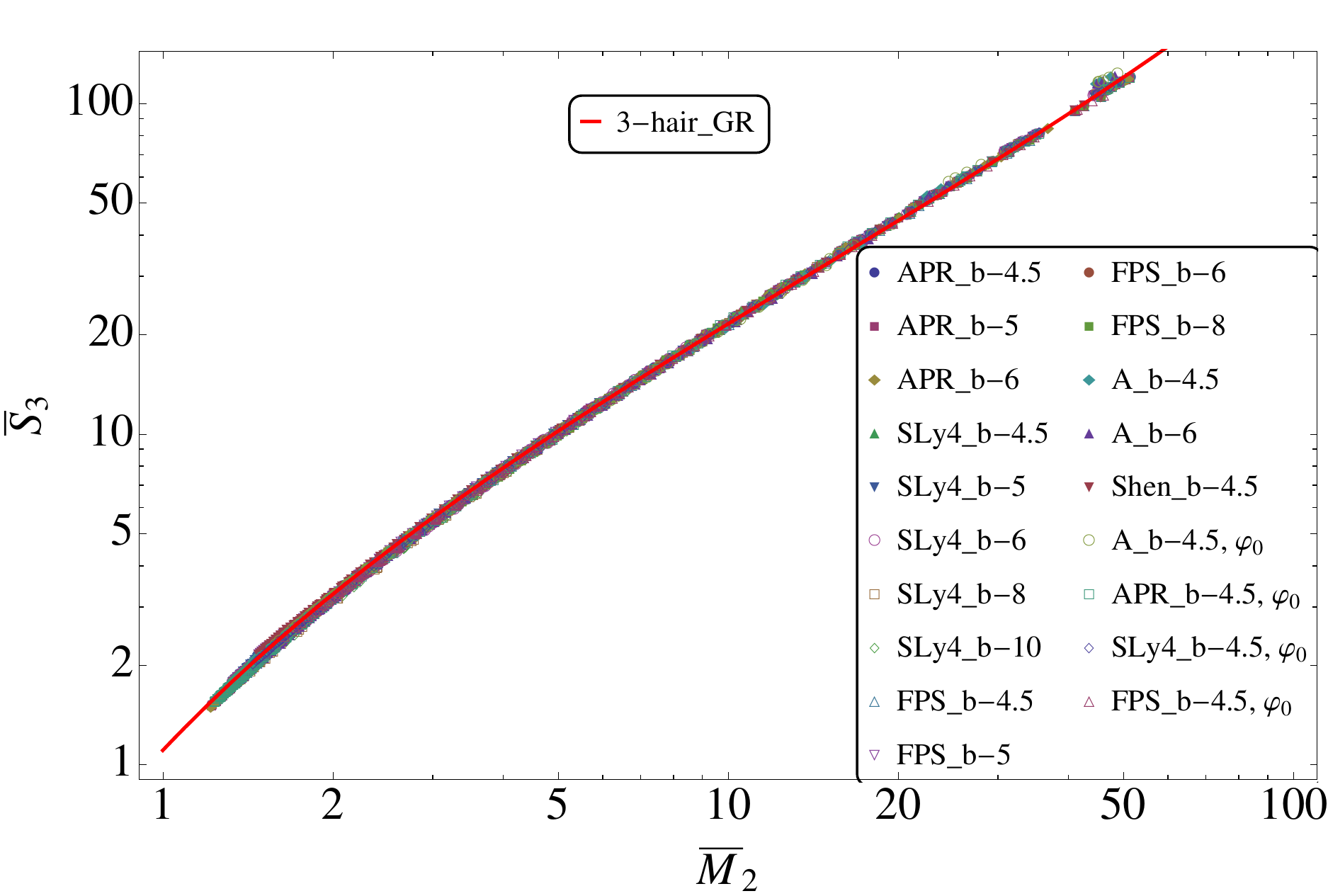}
\includegraphics[width=0.48\textwidth]{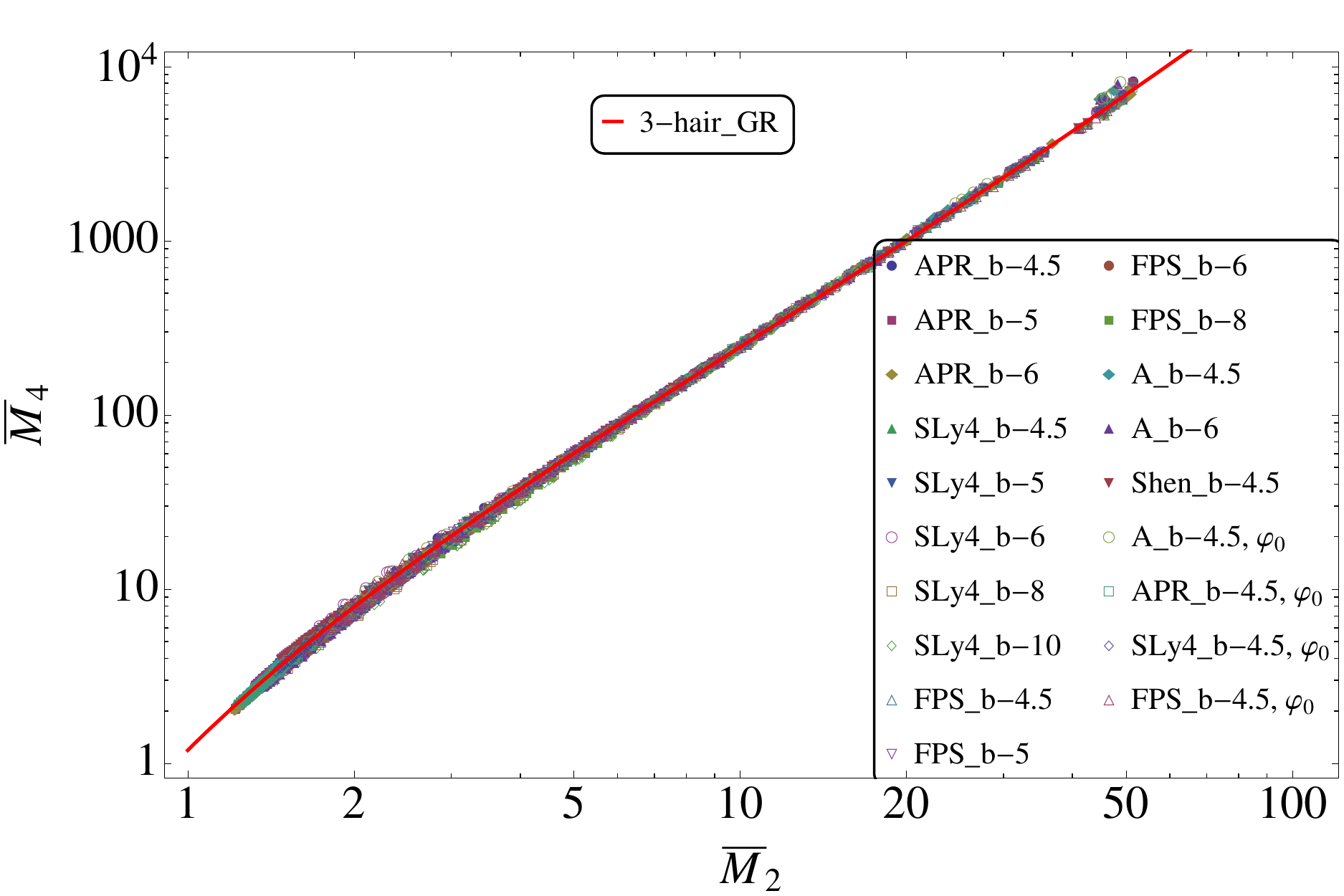}
\caption{\emph{3-hair relations between mass and angular momentum moments.}
The figure on the left shows the relation between the spin octupole and the quadrupole while the one on the right shows the relation between the mass hexadecapole and the quadrupole. The data points correspond to scalarized models for various EoSs with $\beta=-4.5,-5,-6,-8,-10$, as well as models for various EoSs with $\beta=-4.5$ and a non-zero asymptotic value of the scalar field $\varphi_0$. On top of the data points we have plotted the GR 3-hair relations as solid red curves. As one can see, independent of the theory, all the points trace the GR curves. Therefore, the 3-hair relations are the same in ST theory as in GR. The quantities plotted are the reduced moments, i.e., $\bar{M}_2 \equiv -M_2/(j^2M^3)$, $\bar{S}_3 \equiv -S_3/(j^3M^4)$, and $\bar{M}_4 \equiv M_4/(j^4M^5)$, as they are defined in scalar-tensor theory in the Einstein frame \cite{Pappas:2015moments}. 
}
\label{fig:3-hair}
\end{figure}

We now proceed with the presentation of our results. Our first results concern the mass and angular momentum moments of scalarized stars and their behaviour with respect to their GR counterparts. These results are shown in Fig.~\ref{fig:3-hair}, where we have plotted $\bar{S}_3$ against $\bar{M}_2$ and $\bar{M}_4$ against $\bar{M}_2$. The figures include models within the full range of the $\beta$ parameter that we have used  both zero and non-zero asymptotic values of the scalar field. 

As shown in Fig.~\ref{fig:3-hair}, the $\bar{S}_3-\bar{M}_2$ and $\bar{M}_4-\bar{M}_2$ relations of GR \cite{Pappas2014PhRvL,Yagi2014PhRvD} hold for scalarized stars as well. It would be useful at this point to contemplate on this very interesting result. 
Considering the  Einstein frame multipole moments as GR moments with some additional corrections due to the scalar field, one might be tempted to conclude that this results is expected. Indeed, in certain cases it has been argued  
\cite{Sotani:2004rq} 
that the main effect of the scalar field is to stiffen the Einstein frame EOS  with respect to the prescribed Jordan frame EOS [c.f. eqs.~\eqref{EOSconf}].
However, the presence of the scalar field is not in general trivially equivalent to an EOS change, since the gradient of the scalar field itself also acts as a source in the field equations. Furthermore recent studies on $I-Q$ relations for scalarized stars have shown, in contrast to what we find here for the 3-hair relations, that for large values of $|\beta|$ the scalarized $I-Q$ relations can somewhat deviate from the corresponding GR $I-Q$ relations \cite{Pani:2014jra,Doneva2014}. Therefore, what we find here for the 3-hair relations and for values of $\beta$ as much as $\beta=-10$ is quite intriguing. Overall it seems that the 3-hair universal relations are quite less sensitive to the choice of $\beta$ than the $I-Q$ universal relations, being in a sense more universal with respect to different theories of gravity.

The results presented here though, do not eliminate the possibility that stars with an extremely high degree of scalarization in the context of STT or in the context of exotic object in GR ({\em e.g.}~a mixed boson-neutron star) could deviate from these relations. Such objects are beyond the scope of this investigation.

\begin{figure}[h]
\includegraphics[width=0.49\textwidth]{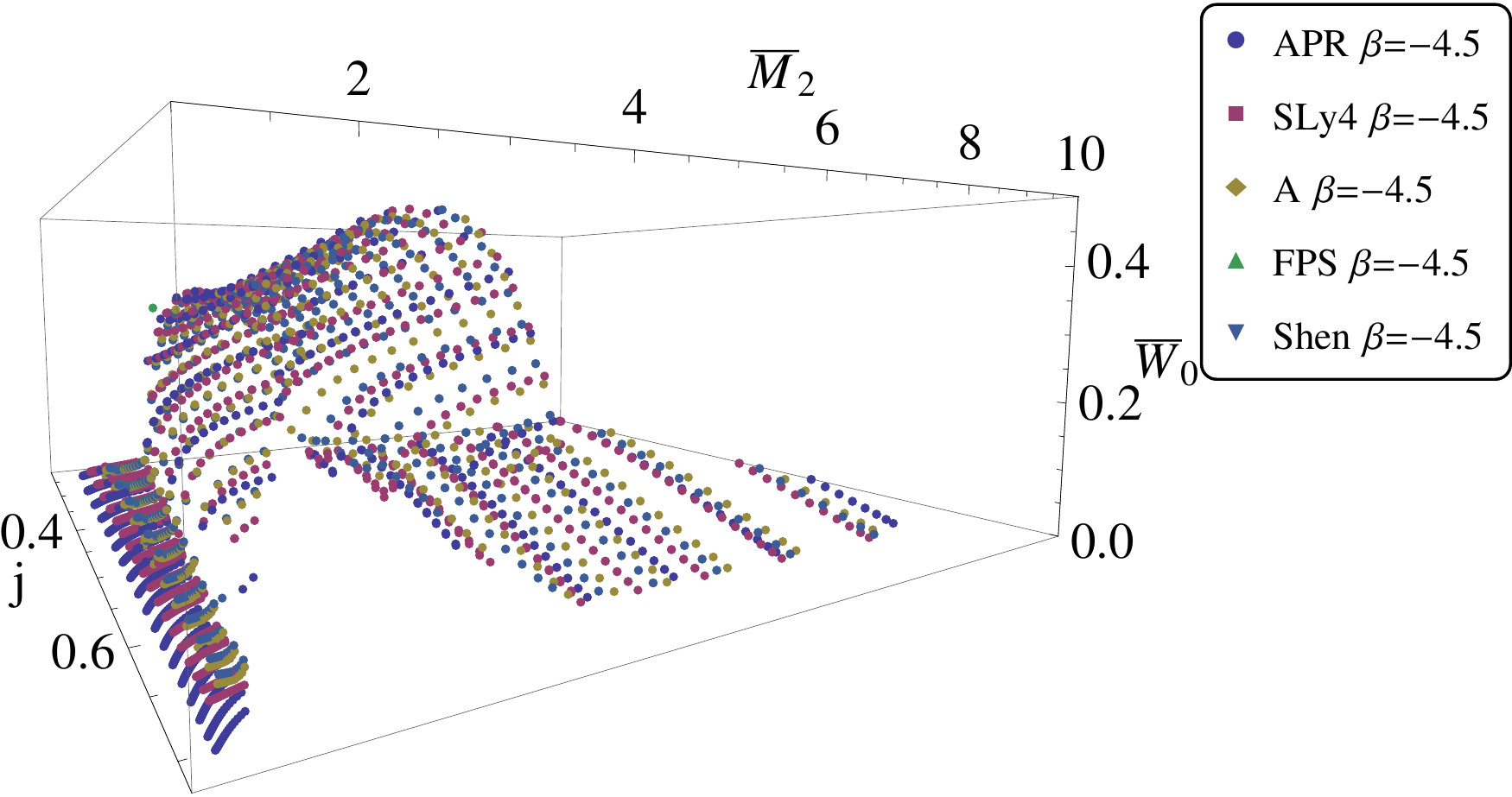}
\includegraphics[width=0.49\textwidth]{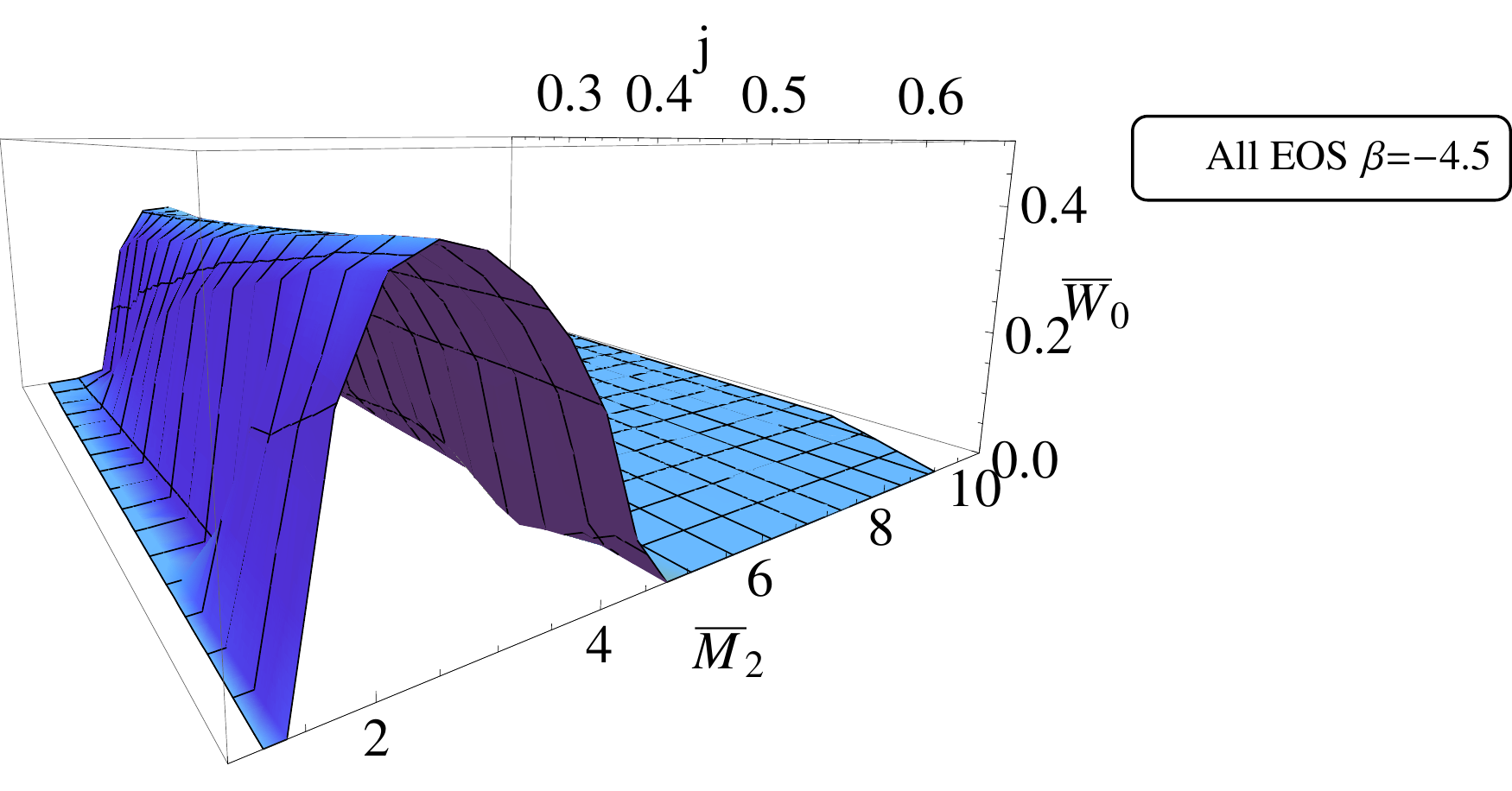}
\includegraphics[width=0.49\textwidth]{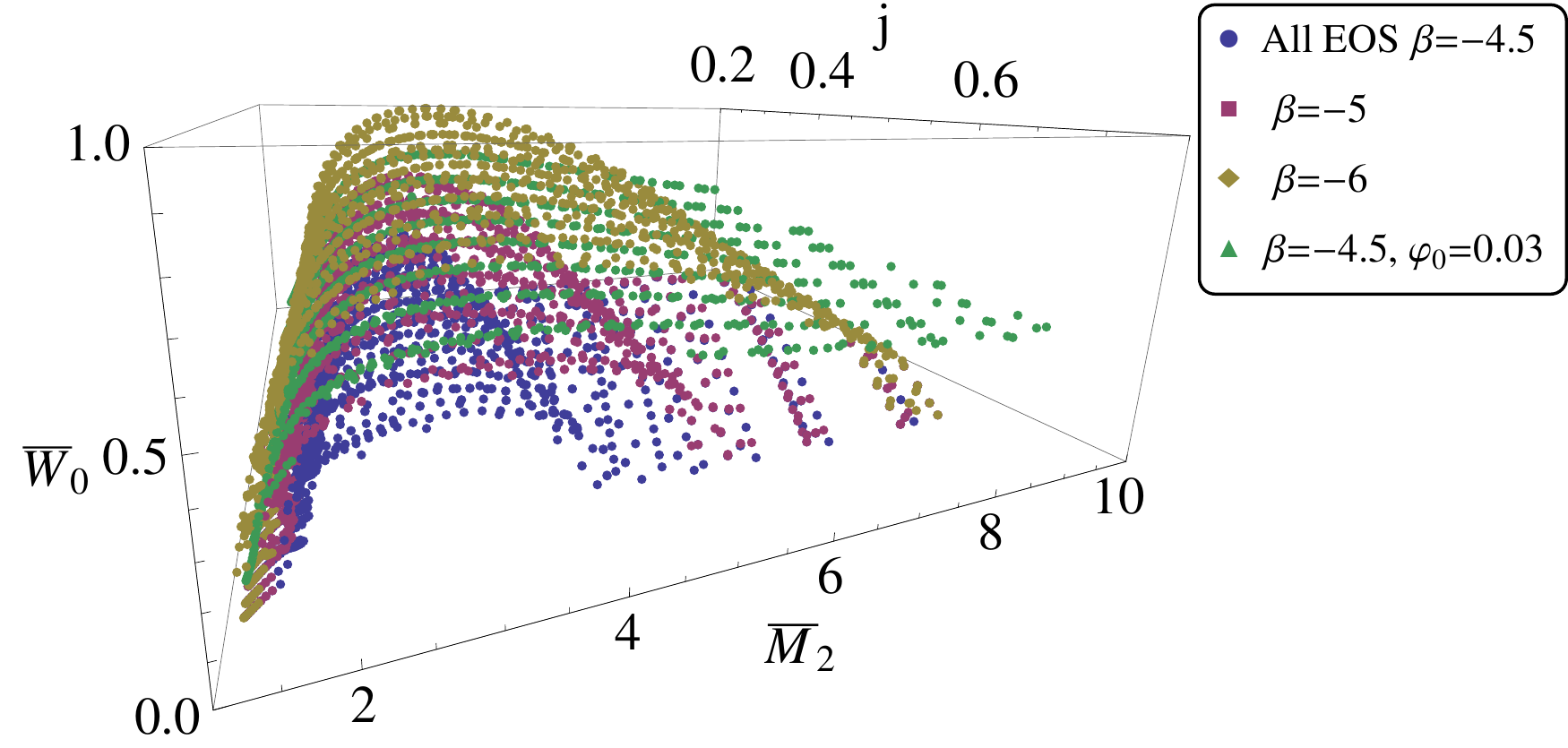}
\caption{\emph{Scalar charge.}
The plots in this figure show the scalar monopole $W_0$ as a function of the spin parameter $j\equiv J/M^2$ and the reduced quadrupole $\bar{M}_2 \equiv -M_2/(j^2M^3)$. The scalar charge demonstrates a universal behaviour, i.e., all the EoSs form the same surface. The relevant surface though, changes depending on the value of $\beta$ of the theory and the asymptotic value of the scalar field $\phi_0$. The quantity that is plotted is the reduced scalar charge $\bar{W}_0\equiv -W_0/(j^a M)$, where $a=0.3$. The reason that this scaling that includes the spin parameter was chosen is because the degree of scalarization has also a spin dependance, so the idea was to try to flatten out the effect. The same scaling with respect to $j$ seems to work for all the different theory cases. The two upper plots correspond to exactly the same data (the middle plot is the surface formed by the points of the top plot). 
}
\label{fig:w0}
\end{figure}

Having seen how the mass and angular momentum moments behave we turn to the scalar field and its moments. The first quantity of interest is the scalar charge or scalar monopole $W_0$. The degree of scalarization of a neutron star will depend on the value that we choose for the parameter $\beta$, with more negative values leading to more scalarized stars and therefore larger values of the scalar monopole as well as the higher order scalar moments. In Fig.~\ref{fig:w0} we show the reduced scalar charge $\bar{W}_0\equiv -W_0/(j^a M)$ as a function of $j$ and $\bar{M}_2$ for models with $\beta=-4.5, -5, -6$ and $\varphi_0=0$ (top plots), as well as a model with $\beta=-4.5$ and $\varphi_0=0.03$ (bottom plots). 

We should recall at this point some general properties of the models that will help the reader interpret the plots. As we mention above, the scalar moments are given in terms of $j$ and $\bar{M}_2$. Increasing value of $j$ corresponds to increasing rotation rate of the star and the higher the degree of scalarization the higher the maximum spin that the models can have. Neutron stars in GR can have a spin up to $j_{\rm max}=0.7$ independent of the EOS (see \cite{Lo2011ApJ,Pappas2015MNRAS} for more details) but scalarized stars can have larger spins. Larger values of $\bar{M}_2$ correspond to less compact objects of lower mass (values larger than 10 usually correspond to masses around or less than $1M_{\odot}$), while the more compact objects with masses close to the maximum mass have the smallest value of $\bar{M}_2$, which tends to $1$. 
The plots show that for large $\bar{M}_2$ the models are not scalarized, while, as $\bar{M}_2$ decreases, at some point stars are spontaneously scalarized and the scalar monopole becomes non-zero. Eventually at small enough values of $\bar{M}_2$, the models become unscalarized again. The degree of scalarization of a neutron star is not independent of the spin. More rapidly rotating neutron stars tend to be more scalarized. To counter this effect to some degree we have chosen to normalise the scalar monopole as $\bar{W}_0\equiv -W_0/(j^a M)$, where the spin weight is $a=0.3$. The spin normalisation was introduced initially in the hope of eliminating the spin dependence, but this has not been possible for any value of $a$.  In spite of this, we have decided to keep this normalisation for all the moments in order to minimise their variation  due to the spin. This point will be further discussed when it arises again.

Fig.~\ref{fig:w0} shows that, within the same theory, i.e., for the same value of $\beta$, all models fall on the same surface independent of the EOS, which means that the scalarized monopole has a universal behaviour. For different theories (different $\beta$s), or for different asymptotic values of the scalar field, the surfaces are different. Unfortunately, the surfaces shown in Fig.~\ref{fig:w0} are not  easy to fit with some simple function. Spontaneous scalarization is a phase transition that occurs at a threshold and finding some empirical relation that would express this threshold in terms of the moments is not straightforward.  One last thing to note is that in the case where the asymptotic value of the scalar field is not zero, the models are scalarized even for small compactnesses (or large $\bar{M}_2$) as we can see in the bottom plot of Fig.~\ref{fig:w0}.     
%

\begin{figure}[h]
\includegraphics[width=0.49\textwidth]{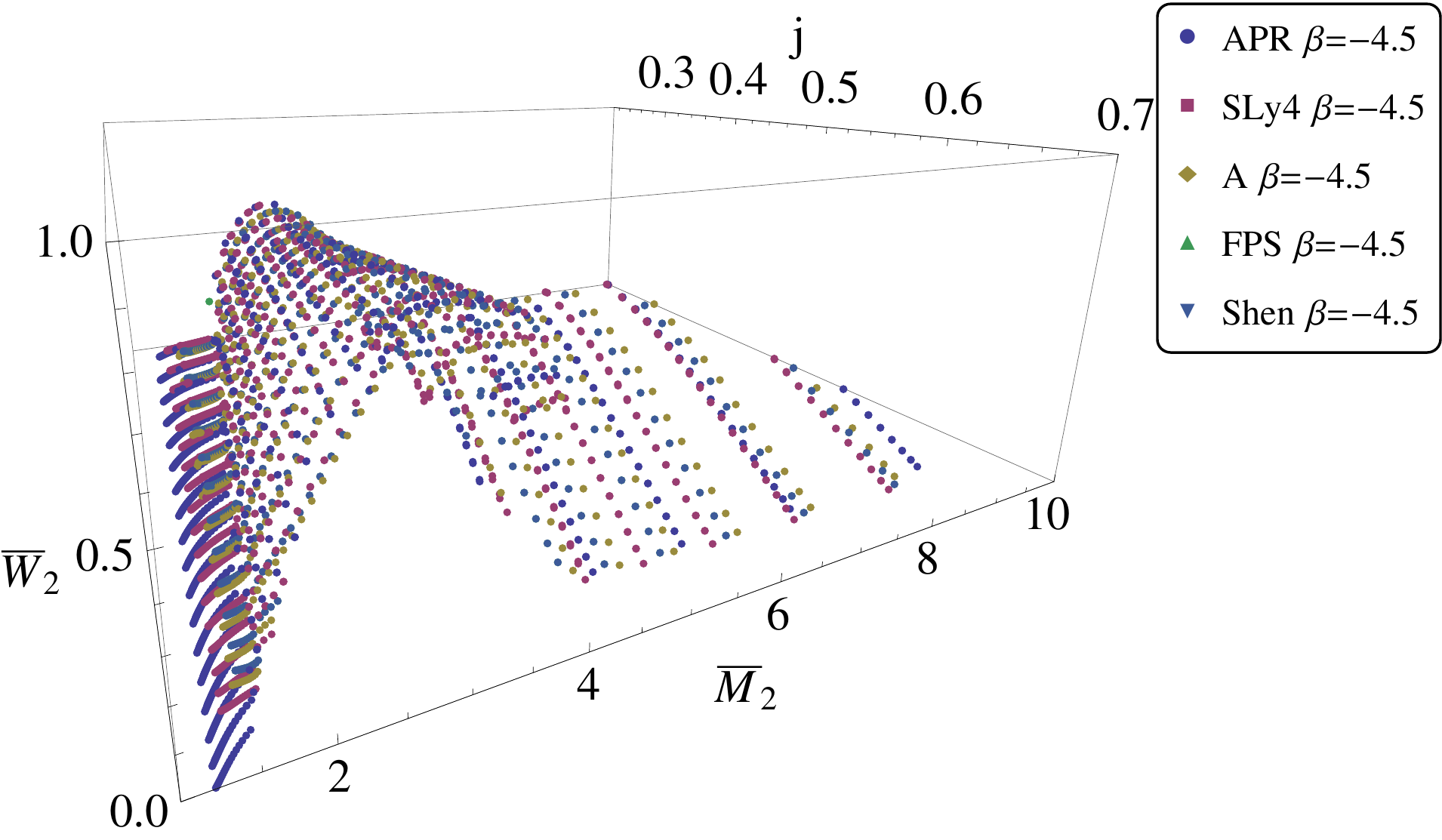}
\includegraphics[width=0.49\textwidth]{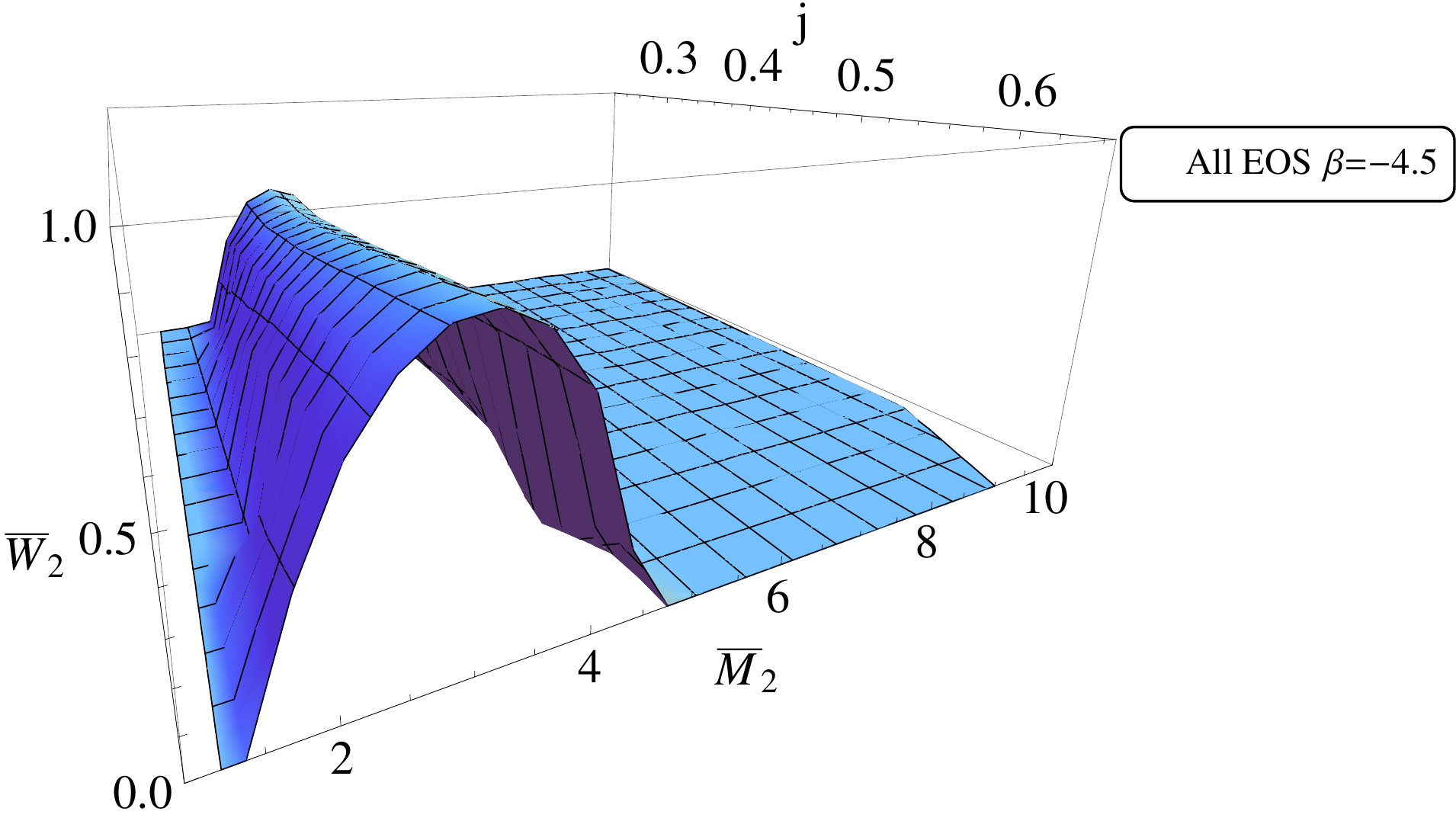}
\includegraphics[width=0.49\textwidth]{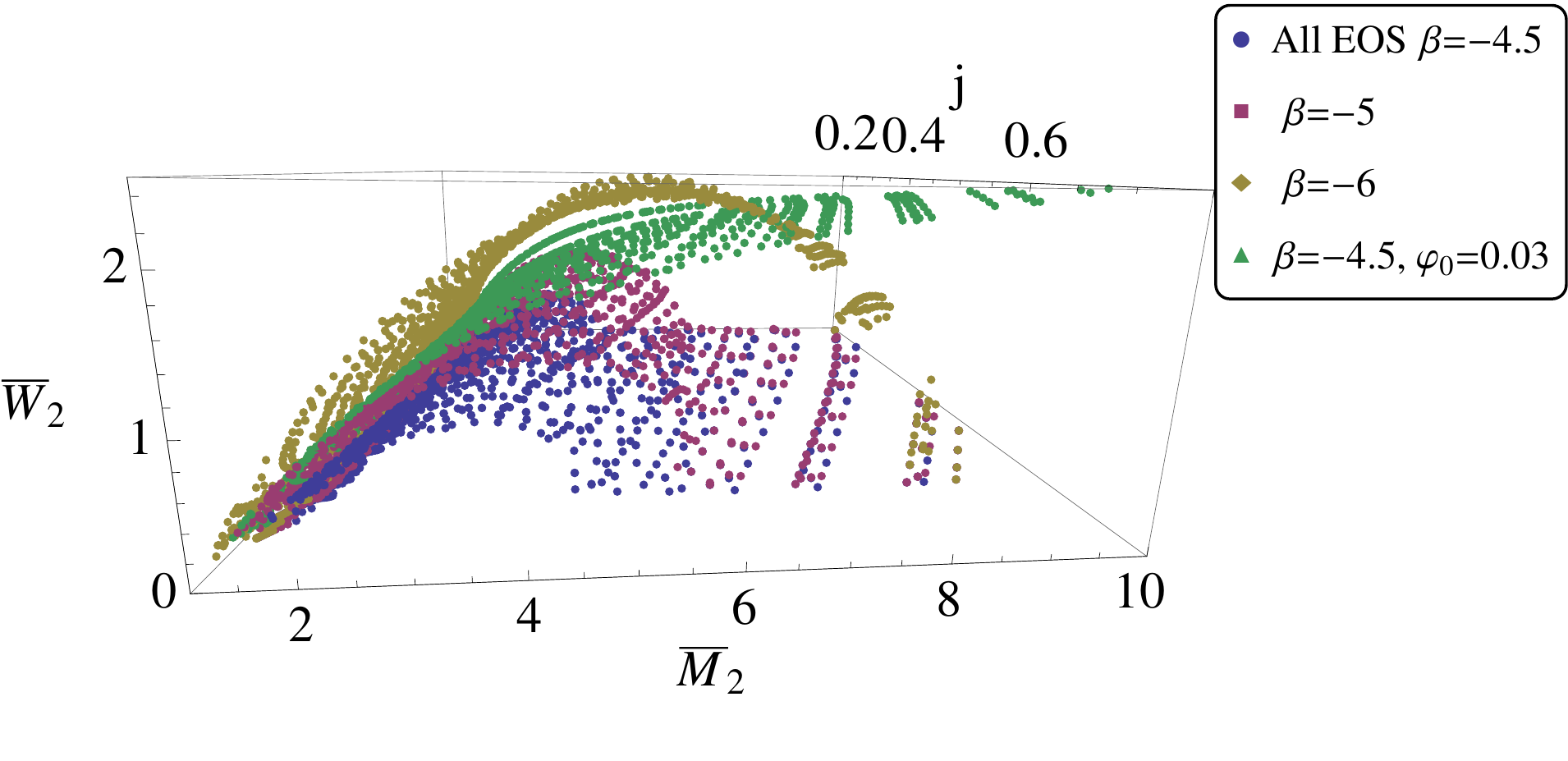}
\caption{\emph{Scalar quadrupole.}
The plots in this figure show the scalar quadrupole $W_2$ as a function of the spin parameter $j=J/M^2$ and the reduced quadrupole $\bar{M}_2=-M_2/(j^2M^3)$. Same as in the previous figure we observe universal behaviour. The quantity that is plotted is the reduced scalar charge $\bar{W}_2\equiv W_2/(j^a M^3)$, where $a=5/3$.}
\label{fig:w2}
\end{figure}

We now turn our attention to the next scalar moment, the scalar quadrupole. The behaviour of the reduced scalar quadrupole $\bar{W}_2\equiv W_2/(j^a M^3)$ is similar to what we saw for the scalar monopole and is presented in Fig.~\ref{fig:w2}. As for the monopole, we have assigned a spin weight to the normalisation of $W_2$ which is $a=5/3$. One could assume that the scalar quadrupole would be driven by the mass quadrupole of the star and therefore the spin dependance would be $\sim j^2$, but as it turns out, the behaviour is more complicated than that. For this reason we have chosen to normalise the multipole in this way in order to reduce the variation due to the spin,  as we did for the scalar monopole,. Similarly to the monopole, different choices of $\beta$ and $\varphi_0$ correspond to different surfaces in the parameter space, while all EOSs for the same theory fall on the same surface.

\begin{figure}[h]
\includegraphics[width=0.49\textwidth]{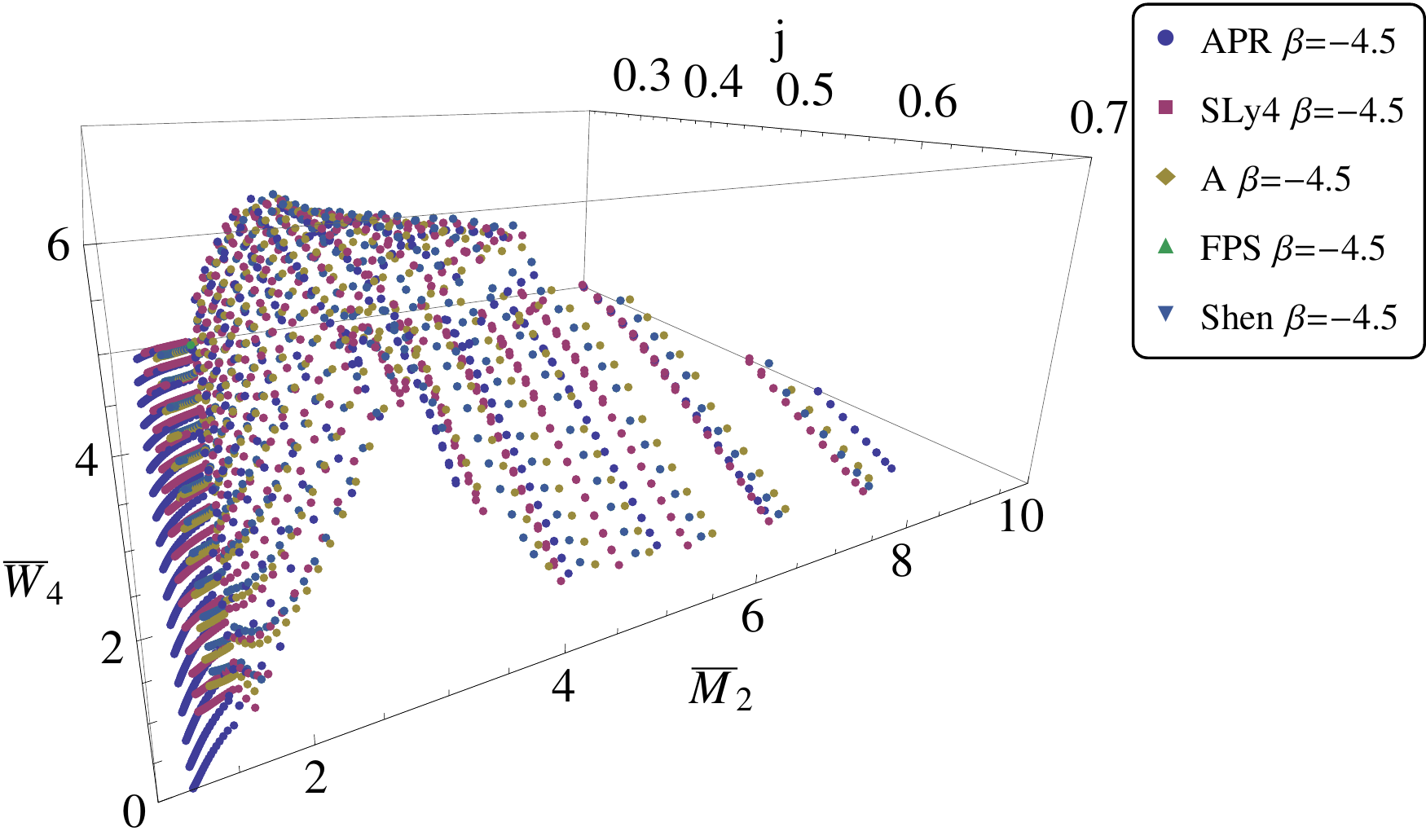}
\includegraphics[width=0.49\textwidth]{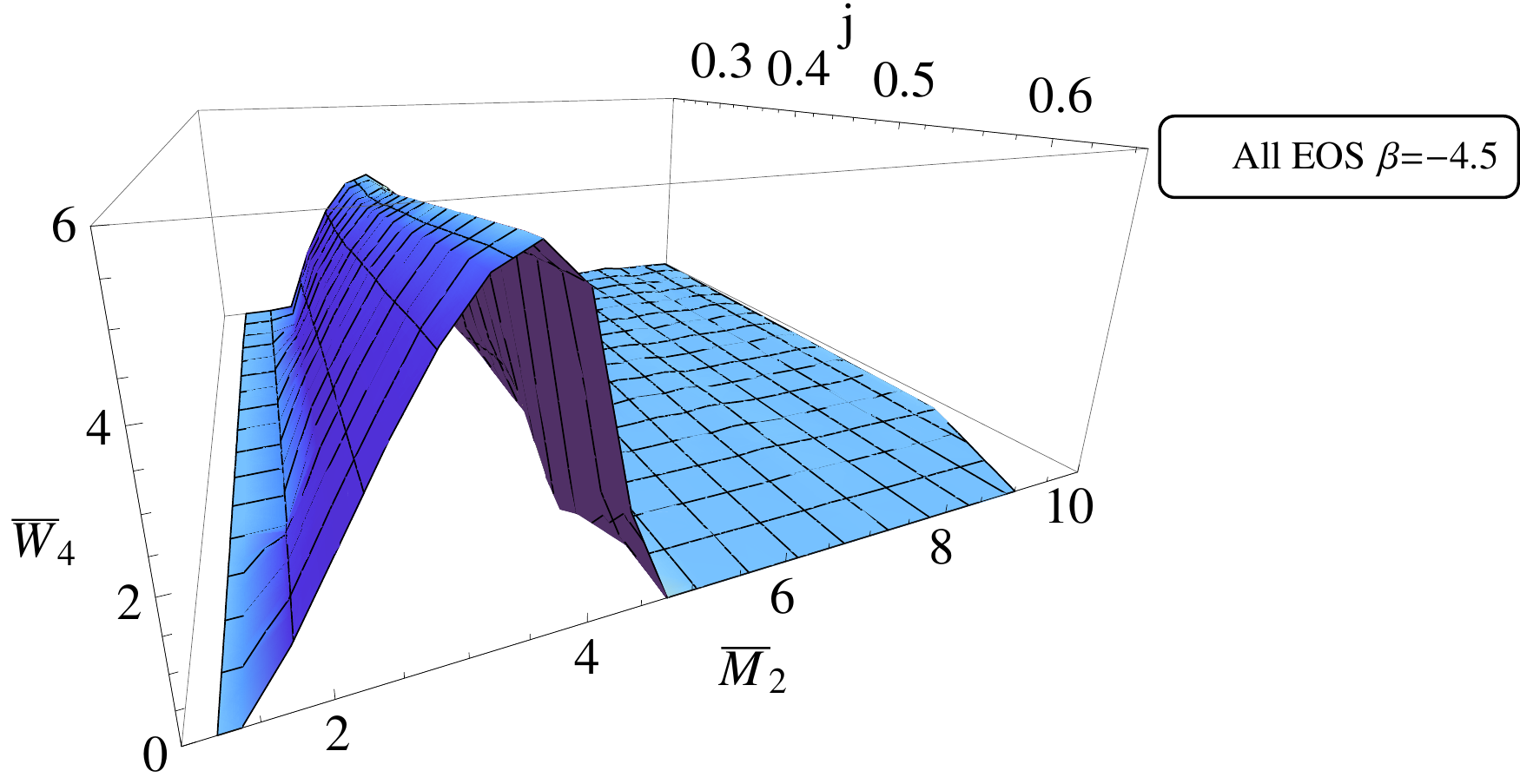}
\includegraphics[width=0.49\textwidth]{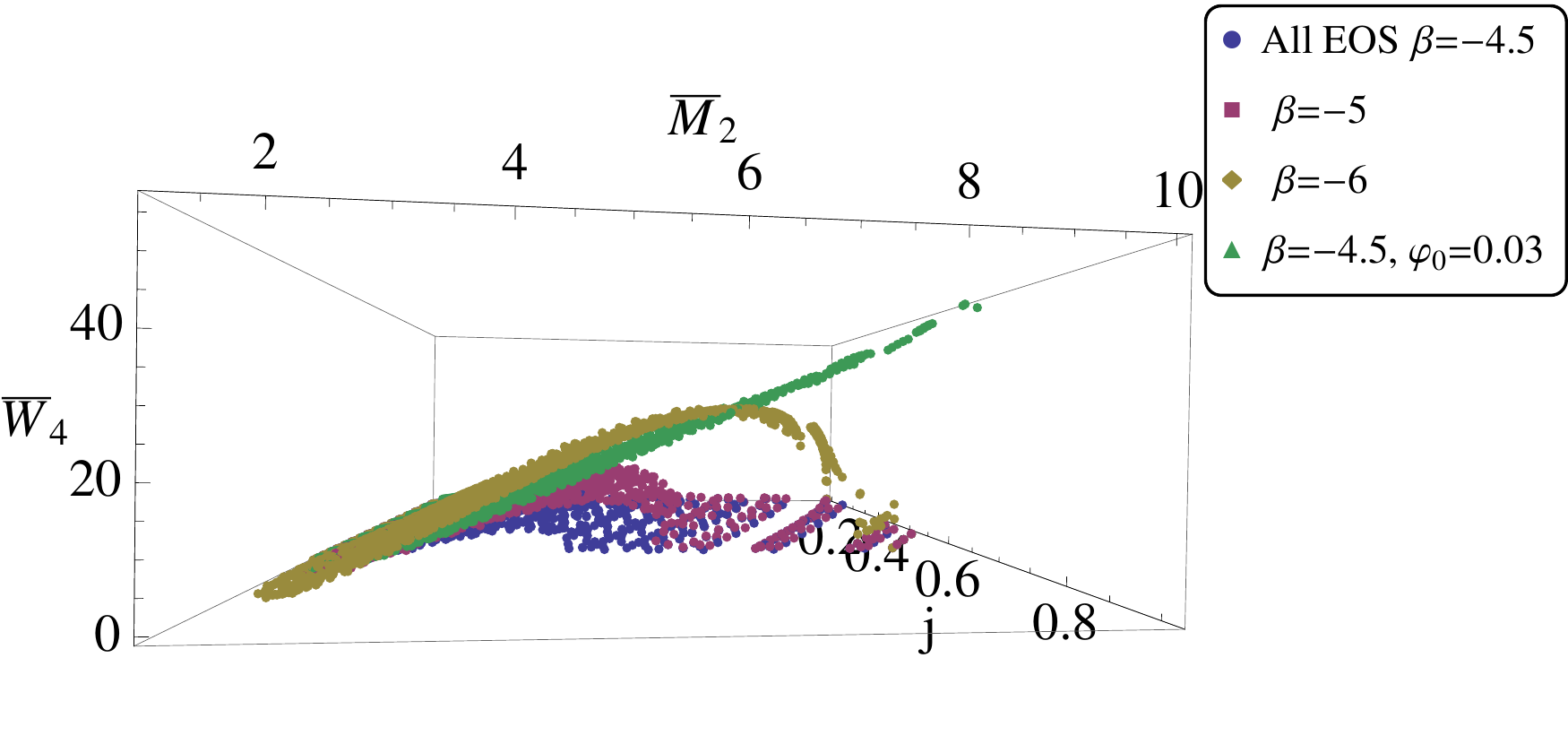}
\caption{\emph{Scalar hexadecapole.}
The plots in this figure show the scalar hexadecapole $W_4$ as a function of the spin parameter $j=J/M^2$ and the reduced quadrupole $\bar{M}_2=-M_2/(j^2M^3)$. Same as in the previous figures we observe universal behaviour. The quantity that is plotted is the reduced scalar charge $\bar{W}_4\equiv -W_4/(j^a M^5)$, where $a=3.6$.}
\label{fig:w4}
\end{figure}

The last scalar moment that we have calculated from the numerical models is the scalar hexadecapole $W_4$. The results for the reduced scalar hexadecapole $\bar{W}_4\equiv -W_4/(j^a M^5)$, where $a=3.6$, are given in Fig.~\ref{fig:w4}. Again we observe a behaviour similar to the previous two cases. The bottom line of this analysis is that the scalar moments in the Einstein frame demonstrate an EOS independent behaviour for the same parameters $\beta$ and $\varphi_0$ following the same surfaces in the respective parameter spaces, while as we change $\beta$ the moments fall on clearly separate surfaces. Therefore, while EOS uncertainties can be circumvented just as in GR, as the scalar moments demonstrate universal behaviour, 
one can identify different theories (different $\beta$s) as they correspond to different surfaces for $\bar{W}_0$, $\bar{W}_2$, and $\bar{W}_4$.

\section{Relating moments to observables and comparison to GR.}
\label{sec:observables}

In the previous section we saw how the moments in the Einstein frame exhibit universal behaviour with respect to the different EOSs of nuclear matter. We also saw that in the case of mass and rotation moments the universal relations found in GR also capture the behaviour of the moments of scalarized stars independently of the specific theory chosen. The latter property doesn't hold for the scalar moments. While they are EOS independent within a specific theory, they do depend on the choice of a particular theory 
(reflected on the choice for the value of $\beta$). 
But as we have already mentioned, the Einstein frame moments are not directly observable and if we want to connect our results to astrophysical observations we will have to calculate physical quantities in the Jordan frame. The transformation to the Jordan frame is a conformal transformation of the form, $\tilde{g}_{\mu\nu}= {\cal A}^{2}(\varphi)g_{\mu\nu}$, and we have defined earlier in Section \ref{sec:setup} the coupling function $k(\varphi)= {d\ln({\cal  A}(\varphi))}/ {d\varphi}$. 

It is common in the literature to use the Damour~\&~Esposito-Far\'ese notation for the asymptotic expansion of this quantity, i.e., $k(\varphi)|_{\infty}=\alpha$,  $\left(d k/d\varphi\right)_{\infty}=\beta$, $\left(d^2 k/d\varphi^2\right)_{\infty}=\gamma$, and so on. In our case and for what follows, due to the form of the coupling function that we have been using, $\alpha$, $\gamma$ (in this notation) and all the higher derivatives will be zero. Furthermore, since current constraints point towards a zero asymptotic value of the scalar field we will assume that at infinity we have $\varphi_0=0$. We stress that we have instead used $\alpha$, $\gamma$ to denote metric functions above. 

Returning to the question of observables, the natural choice is to consider observables that are related to the geodesics of the spacetime. Such observables and their connection to the moments both in GR and in STTs are discussed in what follows.    

\subsection{Observables and moments}

It was shown by Ryan \cite{Ryan95} that there are quantities associated to the geodesics of a GR spacetime that can be expressed in terms of expansions where the coefficients depend on the multipole moments. The expansions of these same quantities have also been calculated for STTs with a massless scalar field \cite{PappasMNRAS2015ST}. 
These quantities are : (i) the change of energy per unit mass of a test particle ($\tilde{E}$) per logarithmic orbital frequency interval for equatorial circular geodesics, denoted by $ \Delta \tilde{E}$; (ii)  the ratio of the periastron precession frequency ($\Omega_r$) of a slightly eccentric equatorial orbit over the orbital frequency ($\Omega$) of the corresponding circular orbit, $\Omega_r/\Omega$; (iii) 
the ratio of the nodal precession frequency ($\Omega_z$) of a slightly off-equatorial orbit over the orbital frequency of the corresponding circular equatorial orbit, $\Omega_z/\Omega$. 
The expansion parameter is $U\equiv (\bar{M}\Omega)^{1/3}$, 
which corresponds to the orbital velocity of the test particle. The quantity $\bar{M}\equiv M-\alpha W_0$ corresponds to the Keplerian mass that one would measure from the motion of a companion star, if the system was part of a binary.  

Here, we briefly present these expansions in GR and in scalar-tensor theory, as they were derived in \cite{Ryan95,PappasMNRAS2015ST}, up to the same corresponding order in $U$ and taking into account the constraints we have from the ansatz that we have used. The energy change per logarithmic orbital frequency change in GR up to $O\left(U^{8}\right)$ is given by the expression, 
\begin{eqnarray}
 \Delta \tilde{E}=-\frac{U}{3}\frac{d\tilde{E}}{dU}&=& 
\frac{U^2}{3}-\frac{U^4}{2}+\frac{20 S_1 U^5}{9 M^2} \nn \\
&& + \left(\frac{M_2}{M^3}-\frac{27}{8}\right)U^6 \nn \\
&& +\frac{28 S_1 U^7}{3 M^2}%
+O\left(U^{8}\right),
\end{eqnarray}
while the corresponding expression in scalar-tensor theory is, after setting $\alpha=\gamma=0$ as discussed above, 
\begin{align}
 \Delta \tilde{E}=& \frac{U^2}{3}
 +\left(
 \frac{2 \beta  W_0^2}{9 \bar{M}^2}
 -\frac{1}{2}\right)U^4 
 +\frac{20 S_1 U^5}{9 \bar{M}^2}  \nn \\
 +%
 &\left[
 \left(\frac{M_2}{\bar{M}^3}-\frac{27}{8}\right)+\frac{4 W_0^2
   \bar{M}^2 \left(
   3 \beta +2\right)
   +8   \beta ^2 W_0^4}{24 \bar{M}^4}\right] U^6\nn\\
  & 
  +\frac{28 S_1 U^7 \left(
   9  \bar{M}^2+2 \beta  W_0^2\right)}{27 \bar{M}^4}
+O\left(U^{8}\right),
\end{align}
where we have $\bar{M}= M-\alpha W_0=M$. Similarly, the ratio $\Omega_r/\Omega$ in GR is,
\begin{eqnarray} 
\frac{\Omega_r}{\Omega}&=&3 U^2-\frac{4 S_1 U^3}{M^2}+\left(\frac{9}{2}-\frac{3 M_2}{2 M^3}\right) U^4 \nonumber \\
&&-\frac{10 S_1 U^5}{M^2}%
+O\left(U^6\right),
\end{eqnarray}
while the corresponding expression in scalar-tensor theory is
\begin{eqnarray}  \frac{\Omega_r}{\Omega}&=& \left(3-\frac{\beta W_0^2 %
                                                       }{2 \bar{M}^2}\right)U^2
                                                       -\frac{4 S_1 U^3}{\bar{M}^2}\nn\\
&&+%
\left[
     \left(\frac{9}{2} -\frac{3M_2}{2\bar{M}^3} \right)
   +   \left(
    \beta -1\right) \frac{W_0^2}{\bar{M}^2}
   -\frac{13 \beta ^2 W_0^4}{24 \bar{M}^4}\right] U^4  \nn\\
   &&-\frac{2
   U^5 
   S_1 \left(
   15 \bar{M}^2+5 \beta  W_0^2\right)
   }{3 \bar{M}^4} 
+O\left(U^6\right).
\end{eqnarray}
Finally, the ratio $\Omega_z/\Omega$ in GR is,
\begin{eqnarray}  \frac{\Omega_z}{\Omega}&=&\frac{2 S_1}{M^2}U^3+ \frac{3 M_2 }{2 M^3}U^4 \nn \\
&&+\frac{ \left(3 M M_2+7 S_1^2\right)}{M^4}U^6%
+O\left(U^{7}\right),
\end{eqnarray}
while the corresponding expression in scalar-tensor theory is
\begin{eqnarray}  \frac{\Omega_z}{\Omega}&=&\frac{2 S_1 }{\bar{M}^2}U^3
+\frac{3  %
 M_2 %
}{2 \bar{M}^3}U^4
+\frac{2 S_1 %
\beta  W_0^2%
}{\bar{M}^4}U^5 \nn\\
&&+\frac{U^6 }{2 \bar{M}^5}  [\bar{M} \left(6 M_2 \bar{M}+14 S_1^2\right) \nn \\
&&-3 \beta\bar{M}W_0 W_2%
+5 \beta  W_0^2 %
M_2%
+O\left(U^{7}\right).
\end{eqnarray}

$\Delta \tilde{E}$ is a quantity that is more immediately relevant to gravitational waves and extreme mass ratio inspirals, while the other two quantities can be also relevant to systems such as X-ray binaries, where one observes quasi-periodic oscillations (QPOs) of the X-ray spectrum of the accretion disc around the compact object. If one were to assume, for example, the relativistic precession model for QPOs, by Stella and Vietri \cite{Stella1998ApJ,Stella1999PhRvL}, then one could associate specific QPO frequencies to $\Omega_r$, $\Omega_z$, and $\Omega$.\footnote{There are other models as well, such as some of the models derived from discoseismology, where oscillations of the disc can be associated to the geodesic frequencies \cite{Kato1990PASJ,Perez1997ApJ,Silbergleit2001ApJ,Lai2009MNRAS,Tsang2016ApJ}.} The relevant observations could then be fitted to recover the coefficients of the expansions.   

\subsection{Measuring the scalar charge and $\beta$.}

Inspecting the expansions in GR and the corresponding expansions in STT reveals that it is possible to distinguish between the two theories, either by  comparing the coefficients of the same order between the two theories or by comparing different order coefficients against each other. 
As we saw in the previous subsection one could expand $\Omega_r/\Omega$ and  $\Omega_z/\Omega$ in terms of powers of $\Omega$ as, 
 $      (\Omega_r/\Omega)= \sum C_a \Omega^{a/3},$ and  $      (\Omega_z/\Omega)= \sum F_a \Omega^{a/3}$,   
where the coefficients $C_a$, for example, will be $C_a=\bar{M}^{a/3} f_a(\bar{M},\beta,W_0,S_1,M_2,W_2,\ldots)$. These coefficients could be used to measure the various parameters. The frequencies that are most commonly observed in low mass X-ray binaries (LMXBs) are the two larger ones, i.e., $\Omega$ and $\Omega_r$. 
These are observed as pairs of kHz QPOs, while occasionally one also observes a third low frequency QPO, which is assumed to be $\Omega_z$. Since the most common occurrence is the former one, we will start assuming that only $\Omega_r$ and $\Omega$ are known. We will then explore how far one can go by using either additional information from $\Omega_z$ or the universal behaviour we have described previously.

\subsubsection{Setting up the problem and constraints.}

In GR one could independently measure the mass from the lowest order term in $\Omega_r/\Omega$, since we have that $C_2^{\rm GR}=3M^{2/3}$. In  scalar-tensor theories however that term has additional contributions due to the scalar field and is of the form $ C_2^{\rm STT}=\left(3-\frac{\beta W_0^2}{2 \bar{M}^2}\right)\bar{M}^{2/3}$. One could go around this problem if an independent measurement of the mass $\bar{M}$ were available. For example, since this sort of QPO producing X-ray sources are LMXBs, the mass could be estimater from the Keplerian motion of the companion and the compact object ($\bar{M}$ is the Keplerian mass after all). In that case, the estimation of $ C_2^{\rm STT}$ would provide a measurement of the combination $\beta W_0^2$, but more importantly would immediately tell us that we have a deviation from GR. 
In GR the higher order coefficients would enable us to measure the higher order moments, while along the way we would find coefficients that would serve as consistency checks, such as $C_5^{\rm GR}$ which is a consistency check on the measurement of $S_1$ from $C_3^{\rm GR}$. In  scalar-tensor theories things are a little more complicated. The coefficients $ C_2^{\rm STT}$, $ C_3^{\rm STT}$ and $ C_5^{\rm STT}$ could serve as a consistency check if one knew the mass $\bar{M}$ independently, but they could also be used to determine the mass since if we combine them we can arrive to the expression,
\be  45 C_3^{\rm STT} \bar{M}^{2/3}=10 C_2^{\rm STT} C_3^{\rm STT}+6 C_5^{\rm STT}, \label{eq:massConst.}\ee
which relates the mass to these coefficients. 
Therefore, even for a system where the mass is unknown, one can estimate it as long as one can accurately estimate the coefficients up to $C_5^{\rm STT}$. This then allows to estimate $S_1$ as well from $C_3^{\rm STT}$. Up to this point we have $\bar{M}$, $\beta W_0^2$, and $S_1$. Turning to the coefficient $C_4^{\rm STT}$ that contains $M_2$ we notice that we cannot estimate it independently. We can only estimate it in combination with $W_0$, i.e., $\left[\left(\frac{9}{2}-\frac{3 M_2}{2 \bar{M}^3}\right)-\frac{W_0^2}{\bar{M}^2}\right]$. The problem lies with our inability so far to separate $\beta$ and $W_0$. Aiming to break the degeneracy  between $M_2$ and $W_0$ by using higher order terms seems a difficult task with uncertain conclusion. 
For instance, while the next order term, $C_6^{\rm STT}$,  includes all the relevant terms, it also includes the scalar quadrupole $W_2$ that first appears in the expansion at that order. 

The situation for measuring the multipole moments and the parameters of the particular STT improves dramatically when we have information for both $\Omega_r/\Omega$ and  $\Omega_z/\Omega$ from a specific system. In that case, we can use the same analysis presented for $\Omega_r/\Omega$ to estimate  $\bar{M}$, $\beta W_0^2$, $S_1$, and the combination $\left[\left(\frac{9}{2}-\frac{3 M_2}{2 \bar{M}^3}\right)-\frac{W_0^2}{\bar{M}^2}\right]$, but then from  $\Omega_z/\Omega$ and the coefficient $F_4^{\rm STT}$ one can estimate $M_2$ and break the degeneracy, while the coefficient $F_5^{\rm STT}$, not present in GR, can serve as an independent verification of the deviation from GR, as well as a consistency check up to that order. 
Therefore additional information from more frequencies allows for the breaking of degeneracies and performing more tests on deviations from GR.

\subsubsection{Using universal relations to overcome degeneracies.}

\begin{figure*}[ht]
\includegraphics[width=0.55\textwidth]{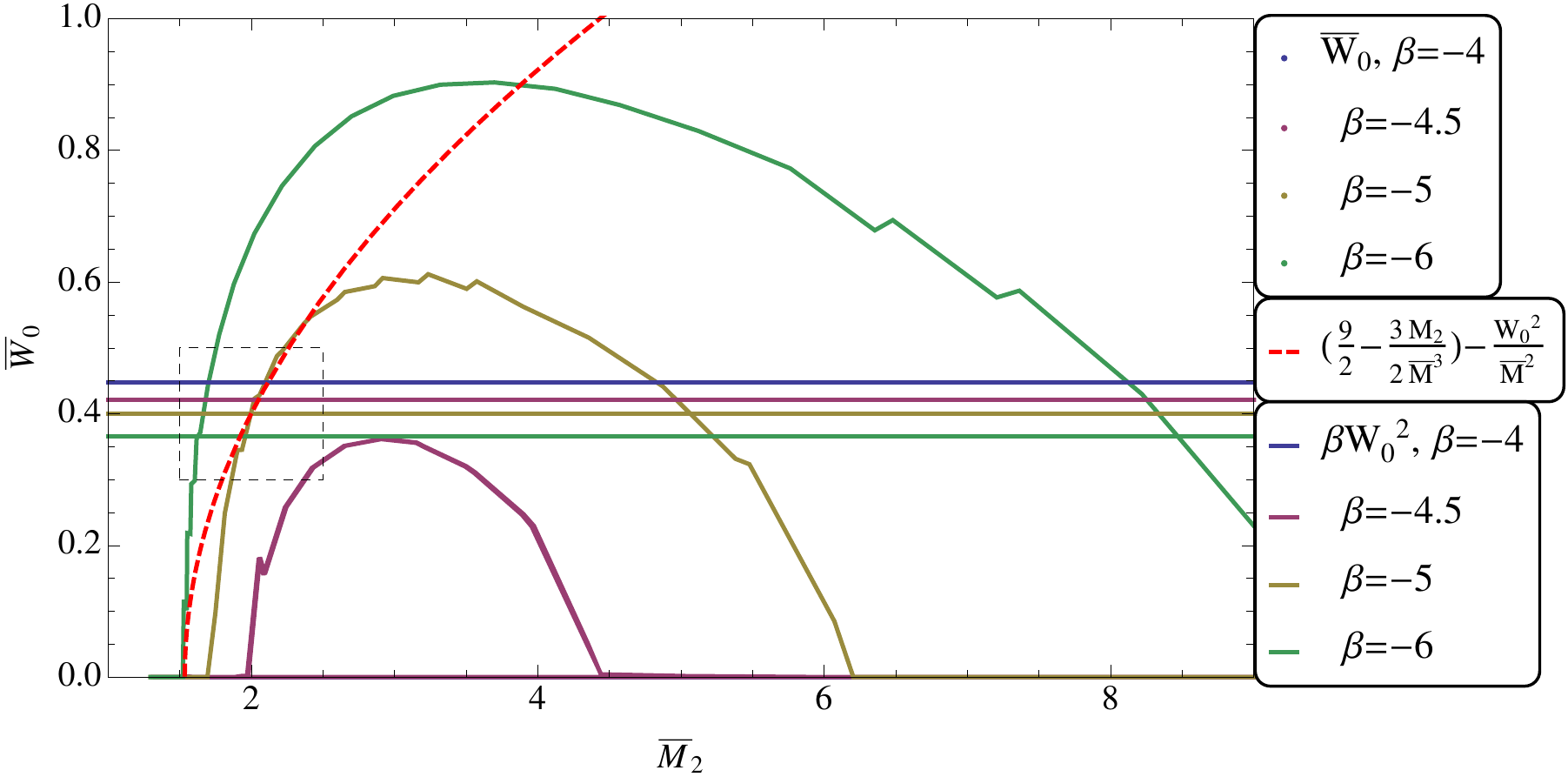}
\includegraphics[width=0.4\textwidth]{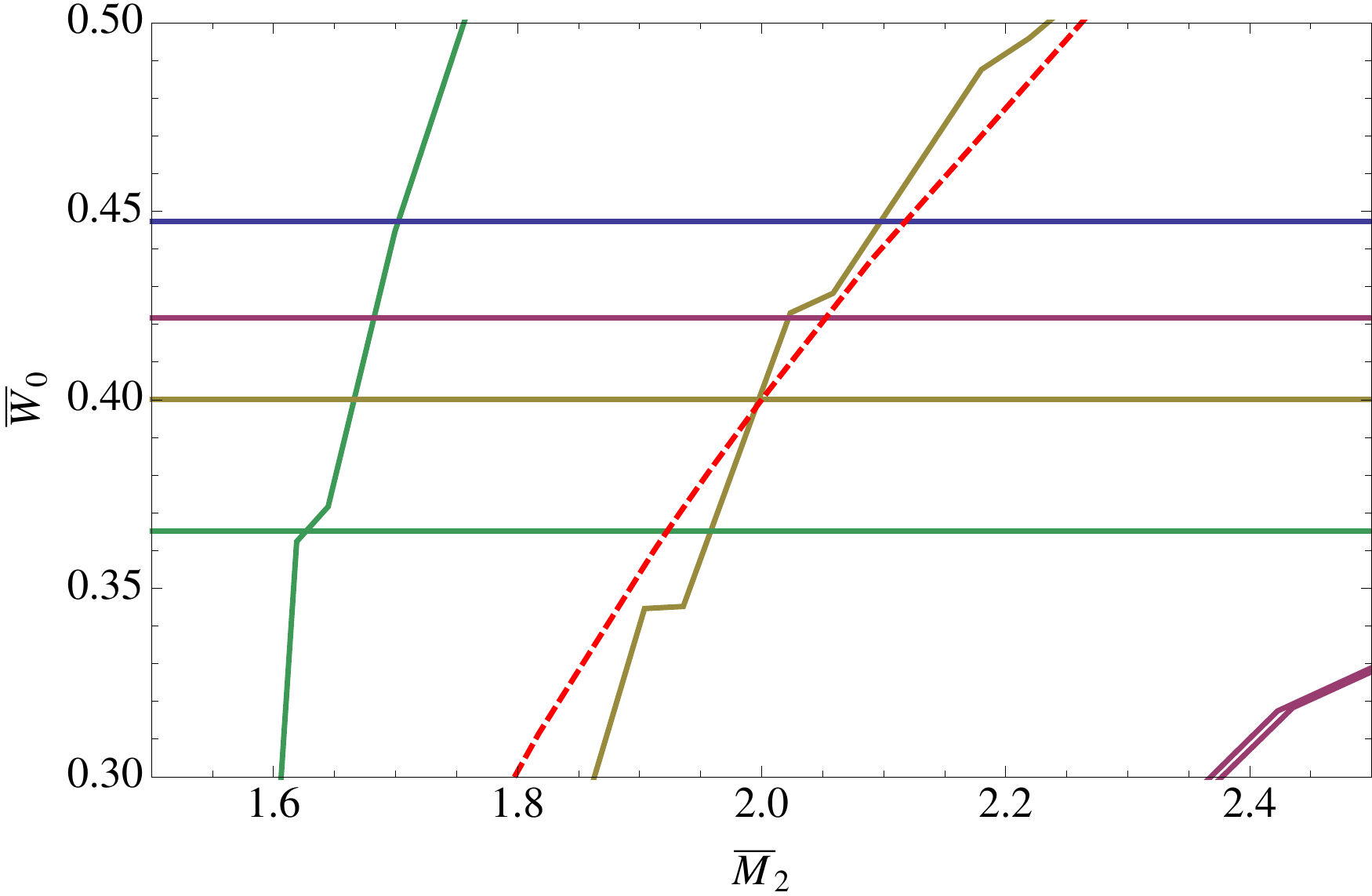}
\caption{\emph{Breaking the $M_2-W_0$ degeneracy.}
The plots in this figure show how one could break the degeneracy between $M_2$ and $W_0$ using the knowledge of the mass $\bar{M}$ (from eq. \ref{eq:massConst.}), the angular momentum $S_1$ (from eq. \ref{qpo:spin}), the constraint for $\beta W_0^2$ from eq. (\ref{w0:constr1}), the constraint from eq. (\ref{w0:constr2}), and the universal relation between $\bar{W}_0$, the spin parameter $j=S_1/M^2$, and the reduced quadrupole $\bar{M}_2$, as it is given in Figure~\ref{fig:w0}. 
The plot on the left shows eq. (\ref{w0:constr1}) for different values of $\beta$ (horizontal lines), the universal relation $\bar{W}_0=f(j,\bar{M}_2)$ for the value of the spin $j$ that has been estimated (in this case $j=0.35$) and for different values of $\beta$, and eq. (\ref{w0:constr2}) which corresponds to the red dashed line. If the results are consistent, then the universal relation cross section curve and the eq. (\ref{w0:constr1}) curve that correspond to the same $\beta$, should intersect with the eq. (\ref{w0:constr2}) curve at the same point. This then indicates the value of $\beta$ and the corresponding values of $\bar{M}_2$ and $\bar{W}_0$. The plot on the right is a magnification of the region where the three curves intersect. The kinks on the curves are due to interpolation errors.}
\label{fig:w0-M2}
\end{figure*}
%
      
In the above discussion we showed that in order to break the $M_2-W_0$ degeneracy one would have to consider both $\Omega_r/\Omega$ and  $\Omega_z/\Omega$, but to reach to that conclusion we did not take into account the results of Section \ref{sec:universal} and the universal behaviour of $W_0/\bar{M}$. 
In fact, if we were to consider only $\Omega_r/\Omega$, the universal behaviour of $W_0/\bar{M}$ and the fact that it can be expressed as some function of $j$ and $M_2$, could be used to break the $M_2-W_0$ degeneracy even without considering $\Omega_z/\Omega$. In what follows we will describe the algorithm that can be used to do this. 

Lets assume that from observations of $\Omega_r$ and $\Omega$ we have estimated the first coefficients of the $\Omega_r/\Omega$ expansion, i.e., $C_2^{\rm STT}$, $C_3^{\rm STT}$, $C_4^{\rm STT}$, and $C_5^{\rm STT}$. The combination of $C_2^{\rm STT}, C_3^{\rm STT}$, and $C_5^{\rm STT}$ will provide the mass $\bar{M}$ of the neutron star, as we describe above. The additional information that we have is that, 

\begin{align} 
                    C_2^{\rm STT}&=\left[3-\frac{\beta}{2}\left(W_0/\bar{M}\right)^2\right]\bar{M}^{2/3},\quad\Rightarrow\quad  \nn \\
                   & \quad\Rightarrow\quad \left(W_0/\bar{M}\right)^2= 2\left(3-C_2^{\rm STT}\bar{M}^{-2/3} \right)\frac{1}{\beta},\label{w0:constr1}\\
                 C_3^{\rm STT}&=   -4 \frac{S_1}{\bar{M}^2} \bar{M}, \quad\Rightarrow\quad j=-\frac{C_3^{\rm STT}}{4\bar{M}},\label{qpo:spin}\\
                 C_4^{\rm STT}&=\left[ \left(\frac{9}{2} -\frac{3M_2}{2\bar{M}^3} \right)-\frac{W_0^2}{\bar{M}^2} \right]\bar{M}^{4/3} \nn \\
                 &+\left( \beta \frac{W_0^2}{\bar{M}^2}
   -\frac{13 \beta ^2 W_0^4}{24 \bar{M}^4}\right) \bar{M}^{4/3}, \quad\Rightarrow\quad \nn \\
   &\quad\Rightarrow\quad \left[ \left(\frac{9}{2} -\frac{3M_2}{2\bar{M}^3} \right)-\frac{W_0^2}{\bar{M}^2} \right]=\bar{C}_4, \label{w0:constr2}
   \end{align}
where, 
\begin{align}
\bar{C}_4 &=\bar{M}^{-4/3}  C_4^{\rm STT}  +\Big[\frac{27}{2}-11 C_2^{\rm STT}\bar{M}^{-2/3} \nn \\
& \;\;\;+\frac{13}{6}\left(C_2^{\rm STT}\right)^2\bar{M}^{-4/3}\Big].
\end{align} 
Eq.\eqref{qpo:spin} straightforwardly gives the spin parameter of the compact object, while Eq. \eqref{w0:constr1} can be interpreted as a bond between $W_0$ and $\beta$ and Eq. \eqref{w0:constr2} relates $W_0$ to $M_2$. Hence, one needs one more bond between these quantities in order to be able to determine them uniquely. The universal relation provide it as follows. First one expresses the equation in terms of the variables of  Section \ref{sec:universal} by using the relations $\left(W_0/\bar{M}\right)^2=j^{0.6}\bar{W}_0^2$ and $(M_2/\bar{M}^3)=-j^2 \bar{M}_2$. One can then effectively consider all quantities as having being uniquely determined, except of $\bar{W}_0$ and $\bar{M}_2$, that instead just satisfy a bond. The additional bond is provided by the universal relation between $\bar{W}_0$ and $(j,\bar{M}_2)$ shown in Figure~\ref{fig:w0}.
 For the estimated value of $j$, the result is a cross section of the surfaces shown in Figure~\ref{fig:w0}, which amounts to having for different values of $\beta$ different curves relating $\bar{W}_0$ to $\bar{M}_2$. 

We can now plot all these constraints on a $\bar{W}_0-\bar{M}_2$ plot, an example of which is shown in Figure~\ref{fig:w0-M2}. The plot shows the $\bar{W}_0=$const. lines that result from the constraint \eqref{w0:constr1} for the different values of $\beta=-4,-4.5,-5,$ and $-6$. It also shows the universal $\bar{W}_0=f(\bar{M}_2)$ curves for the corresponding values of $\beta$ (we note that for the spin in this example, the $\beta=-4$ models are unscalarized). The last curve shown is the constraint resulting from Eq. \eqref{w0:constr2} which is independent of $\beta$ and is therefore a single curve (red dashed curve). In order to have a consistent solution of all of the constraints, the curves of Eq. \eqref{w0:constr1} and Eq. $\bar{W}_0=f(\bar{M}_2)$ that correspond to the same $\beta$ must intersect the curve for Eq. \eqref{w0:constr2} at the same point (or approximately the same point), just as the example in Figure~\ref{fig:w0-M2} shows. From the intersection point one can identify the value of $\beta$ ($=-5$ in this example) as well as the values of $\bar{M}_2$ ($\sim2$) and $\bar{W}_0$ ($=0.4$). 

We have therefore presented an algorithm that makes use of the universal relations for the scalar monopole, in order to measure the parameters of a given scalarized neutron star and the parameters of the corresponding STT from a set of astrophysical observations (pairs of QPO frequencies in this case) that otherwise would not be possible due to degeneracies.


\section{Conclusions}
\label{sec:conclusions}

Universal or EOS independent relations between global properties of neutron stars have proven to be a versatile tool for inferring the properties of neutron stars. These relations have been extensively studied in GR and particular flavours of them, such as the I-Love-Q relations, have been studied in a variety of modifications to GR. The 3-hair universal relations found in GR have been a more difficult problem to tackle, mainly due to the intricacies of defining multipole moments in modifications to GR. 
STT of gravity with a massless scalar field is a class of theories where a definition of moments is already available \cite{Pappas:2015moments}. 

In this work we have computed the multipole moments for scalarized stars in these theories and have shown that they continue to exhibit universal properties. Specifically, we have found that the mass and angular momentum moments follow the same universal 3-hair relations as their GR counterparts \cite{Pappas2014PhRvL,Stein2014ApJ,Yagi2014PhRvD}, independent of the value of the $\beta$ parameter and of the asymptotic value $\varphi_0$ of the scalar field. 
Furthermore we have found that the scalar field moments for every given combination of $\beta$ and $\varphi_0$ exhibit universal behaviour in terms of the spin parameter $j$ and the reduced quadrupole $\bar{M}_2$. That is, when each moment is plotted in terms of $j, \bar{M}_2$ it falls on the same surface independent of the EOS. In addition, different values of $\beta$ and $\varphi_0$ result in different surfaces in the three dimensional space formed by each scalar moment and the two parameters $j, \bar{M}_2$. This appears to be related with the known fact that the degree of scalarizations depends on both the asymptotic value of the scalar field and the value of $\beta$.

Our results demonstrate that the degree of scalarization can be expressed in an EOS independent way, which is still quite intriguing and potentially very useful. In particular, we demonstrate how one can use the universal relations presented here to infer the various properties (i.e., the moments) of a scalarized neutron star, as well as the parameters of the specific STT (i.e., the value of $\beta$), from astrophysical observations.

The algorithm for doing so using LMXBs can be seen as a proof of principle.  A more thorough analysis is necessary in order to determine how accurately the various parameters can be measured and what sort of constrains can be set on STTs from observations. Furthermore, it would be worth exploring how the results presented here could be used in other settings, such as the observation of gravitational waves from the inspiral of NS-NS binaries or BH-NS binaries.

 STT with a massless scalar is only a first step in studying the 3-hair relations in these theories. 
 The next would be to consider STTs with a massive scalar field. This is more challenging, as the multipole moments cannot be defined in the same way as in the massless case. 
 Nevertheless some cases can be easier to handle than others. For instance, if the scalar field were to be fully confined inside the neutron star on account of its large mass, then in the exterior for all practical purposes one would only have to deal with the spacetime, and the moments would be calculated in the same way as in GR. This is something that we will explore in future work.

\begin{acknowledgments}
 
The research leading to these results has received funding from the European Research Council under the European Union's Seventh Framework Programme (FP7/2007-2013) / ERC Grant Agreement n. 306425 ``Challenging General Relativity''. GP acknowledges financial support provided under the European Union's H2020 ERC, Starting Grant agreement no. DarkGRA-757480.  DD would like to thank the European Social Fund, the Ministry of Science, Research and the Arts Baden-W\"urttemberg for the support. DD is indebted to the Baden-Württemberg Stiftung for the financial support of this research project by the Eliteprogramme for Postdocs.   TPS acknowledges partial support from
the STFC Consolidated Grant No. ST/P000703/1. SY  acknowledges financial support by the Sofia University Grants No 3258/2017 and the Bulgarian NSF Grant DCOST 01/6. Networking support by the COST Actions  CA15117, CA16104 and CA16214  is also gratefully acknowledged.

  \end{acknowledgments}

\appendix

\section{Calculation of the multipole moments}
\label{sec:app2}

The calculation of the equilibrium neutron star solutions is done using a modification of the {\tt RNS} code (see \cite{Stergioulas95} for the original GR version of the {\tt RNS} code while the STT extension can be found in \cite{Doneva:2013qva}) and that is why we will follow the formalism and notations that are standard for the KEH method \cite{Komatsu:1989zz,Cook1992}. The coefficients $B_{2l}$, $\nu_{2n,0}$, $\omega_{2n-1,0}$ and $\Phi_{2n}$ are calculated numerically using integrals of the source functions of the field equations \eqref{eq:DiffEq_gamma}--\eqref{eq:DiffEq_omega},\eqref{eq:DiffEq_phi}. In the present paper we consider the case of a zero mass scalar field and thus these integrals are the same in pure GR and in STTs (when the  Einstein frame is employed). \footnote{An alternative approach would be to calculate these coefficient from the asymptotic behaviour of the metric functions, but numerically this is much more imprecise especially if we want to calculate higher order multipole moments.}

First we describe the calculation of the  coefficients $B_{2l}$. The function $B$ in the metric ansatz \eqref{eq:metric_MultipoleMoments} is connected to the metric function $\gamma$ used by the {\tt RNS} code \eqref{eq:metric_RNS} via the relation $B=e^{\gamma}$. Using the integral representation of $\gamma$ \cite{Komatsu:1989zz,Cook1992}, we have
\begin{widetext}
 \begin{align}
\gamma(s,\mu) =& -\frac{2e^{-\gamma/2}}{\pi}  \sum_{n=1}^{\infty} \frac{\sin[(2n-1)\theta]}{(2n-1)\sin\theta}
                           \left[\left(\frac{1-s}{s}\right)^{2n}\int_0^s\frac{ds's'^{2n-1}}{(1-s')^{2n+1}}\int_0^1 d\mu' \sin[(2n-1)\theta']\tilde{S}_{\gamma}(s',\mu')\right.\nn\\
                                        &           +\left.  \left(\frac{s}{1-s}\right)^{2n-2}\int_s^1\frac{ds'(1-s')^{2n-3}}{s'^{2n-1}}\int_0^1 d\mu' \sin[(2n-1)\theta']\tilde{S}_{\gamma}(s',\mu')\right], \label{metricgamma}
\end{align}
with the asymptotic expansion at infinity being 
 \begin{align}
\gamma(s,\mu) =& -\frac{2e^{-\gamma/2}}{\pi}  \sum_{n=1}^{\infty} \frac{\sin[(2n-1)\theta]}{(2n-1)\sin\theta}
                           \left[\left(\frac{1-s}{s}\right)^{2n}\int_0^1\frac{ds's'^{2n-1}}{(1-s')^{2n+1}}\int_0^1 d\mu' \sin[(2n-1)\theta']\tilde{S}_{\gamma}(s',\mu') 
                                     \right], \label{metricgammaasy}
\end{align}
\end{widetext}
where $s$ is the compacted radial coordinate $\left(1-s\right)/s=r_{eq}/r$ with $r_{eq}$ being a characteristic length scale that gives the coordinate equatorial radius of the star, and $\mu=\cos{\theta}$. The source term $\tilde{S}_{\gamma}(s',\mu')$ is connected to the right hand side of eq. \eqref{eq:DiffEq_gamma} and is given by 
\begin{eqnarray}
\tilde{S}_{\gamma} &=& r^2 e^{\gamma/2}\Big\{16\pi e^{2\alpha}p+\frac{\gamma}{2}\Big[16\pi e^{2\alpha} p \nn \\ &&-\frac{1}{2}(\partial_r\gamma)^2-\frac{1}{2r^2}(1-\mu^2)(\partial_\mu\gamma)^2\Big]\Big\}.
\end{eqnarray}

For simplicity, the expression \eqref{metricgammaasy} can be written as 
 \begin{align}
\gamma(r,\mu) =& -\frac{2e^{-\gamma/2}}{\pi}  \sum_{n=1}^{\infty} \frac{\sin[(2n-1)\theta]}{(2n-1)\sin\theta} \frac{\Gamma_{2n}}{r^{2n}} \quad 
\end{align}
where
\begin{align}
\Gamma_{2n} =& ~ r_e^{2n}\int_0^1\frac{ds's'^{2n-1}}{(1-s')^{2n+1}}\int_0^1 d\mu' \sin[(2n-1)\theta']\tilde{S}_{\gamma}(s',\mu'). \label{gammacoeff}
\end{align}
We should also note at this point that, 
\be \frac{\sin[(2n-1)\theta]}{\sin\theta}=\sqrt{\frac{\pi}{2}}T_{2(n-1)}^{1/2}(\mu), \quad \textrm{for } \cos\theta\rightarrow \mu.\ee
With these expressions at hand we can express again the function $\gamma$ in a more convenient way as,
\be \gamma(r,\mu) = -e^{-\gamma/2} \left(\frac{2}{\pi}\right)^{1/2} \sum_{n=1}^{\infty} \frac{T_{2(n-1)}^{1/2}(\mu)}{(2n-1)} \frac{\Gamma_{2n}}{r^{2n}},\ee
where we have it in terms of the Gegenbauer polynomials. This representation is useful since it can be easily related to the asymptotic expansion of the function $B$
\begin{eqnarray}
B&=&1+\sum_{l=0}^{\infty}\frac{B_{2l}}{r^{2l+2}}T_{2l}^{1/2}(\mu) \Rightarrow\nn \\
B&=&1+\left(\frac{\pi}{2}\right)^{1/2}\Big[ \frac{B_0}{r^2} T_0^{1/2}(\mu) +\frac{B_2}{r^4}T_2^{1/2}(\mu) \nn \\
&&+\frac{B_4}{r^6}T_4^{1/2}(\mu)+\ldots \Big].    \label{metricasymptot} 
\end{eqnarray}
Using the orthogonality conditions of the Gegenbauer polynomials
\be \int_{-1}^1d\mu (1-\mu^2)^{1/2}T_l^{1/2}(\mu)T_m^{1/2}(\mu)=\delta_{lm},
\ee
to relate the $\Gamma_{2n}$ coefficients to the $B_{2l}$ coefficients, we have that the $B_{2n}$ coefficients are given as 
\bear B_{2n}&=&\lim_{r \to +\infty} \left(\frac{2}{\pi}\right)^{1/2}  r^{2n+2}\nn\\
   && \times \int_{-1}^1d\mu ~ (1-\mu^2)^{1/2}T_{2n}^{1/2}~(e^{\gamma(r,\mu)}-1). \eear
If we take it's asymptotic expansion in terms of $r$ we can easily obtain
\bear B_{2n}&=&\lim_{r \to +\infty} \left(\frac{2}{\pi}\right)  r^{2n+2}  \int_{-1}^1d\mu ~ (1-\mu^2)^{1/2}T_{2n}^{1/2} \nn\\
                     &&\times ~\left(-\sum_{k=1}^{\infty} \frac{T_{2(k-1)}^{1/2}(\mu)}{(2k-1)} \frac{\Gamma_{2k}}{r^{2k}}+\ldots \right).\eear
The only term that will survive the integration is the $k-1=n$ term that will give the result, 
\be B_{2n}=-\left(\frac{2}{\pi}\right) \frac{\Gamma_{2(n+1)}}{(2n+1)} ,\ee
with the $\Gamma_{2l}$ coefficients given by \eqref{gammacoeff}.

The calculation of the rest of the expansion coefficients $\nu_{2n,0}$, $\omega_{2n-1,0}$ and $\Phi_{2n}$ is more straightforward. Thus, in terms of the source functions in the field equations \eqref{eq:DiffEq_gamma}--\eqref{eq:DiffEq_omega} used by the {\tt RNS} code, these coefficients are given by
 \begin{widetext}
 
  \bea \nu_{2\ell,0}&=&-\frac{r_{eq}^{2\ell+1}}{2}\int_0^1 \frac{ds' s'^{2\ell}}{(1-s')^{2\ell+2}} 
 \int_0^1 d\mu' P_{2\ell}(\mu')\tilde{S}_{\tilde{\sigma}}(s',\mu'),\label{RNSq}\\ \nn \\
 \omega_{2\ell-1,0}&=& \frac{r_{eq}^{2\ell}}{2\ell (2 \ell-1)}\int_0^1 \frac{ds' s'^{2\ell}}{(1-s')^{2\ell+2}}
 \int_0^1 d\mu' (1 - \mu'{}^2) \frac{d P_{2\ell-1} (\mu')}{d \mu'} \tilde{S}_{\hat{\omega}}(s',\mu'),\label{RNSw} \\ \nn \\
 \Phi_{2n}&=&-r_{eq}^{2n+1}\int_0^1 \frac{ds' s'^{2n}}{(1-s')^{2n+2}} 
 \int_0^1 d\mu' P_{2n}(\mu')\tilde{S}_{\phi}(s',\mu'),\label{RNSphi}                           
  \eea
   \end{widetext}
 In the first integral $\ell\geq 0$ while in the second integral $\ell\geq 1$.  The source functions that appear in these integrals are of course connected to the left hand side of the field equations \eqref{eq:DiffEq_sigma}, \eqref{eq:DiffEq_omega}, and \eqref{eq:DiffEq_phi} respectively, i.e., 
  \begin{widetext}
 \begin{align}
 \frac{\tilde{S}_{\tilde{\sigma}}}{r^2}(r,\mu)=&e^{\gamma/2}\left[8\pi e^{2\alpha}(\varepsilon+p)\frac{1+v^2}{1-v^2}+r^2(1-\mu^2)e^{-2\tilde{\sigma}}  \left[(\partial_r \omega)^2+\frac{1}{r^2}(1-\mu^2)(\partial_\mu\omega)^2\right] 
 +\frac{1}{r}\partial_r\gamma-\frac{1}{r^2}\mu \partial_\mu\gamma \right.\nn\\
 &+\left.\frac{\tilde{\sigma}}{2}\left\{16\pi e^{2\alpha}-\partial_r\gamma \left(\frac{1}{2}\partial_r\gamma+\frac{1}{r}\right)-\frac{1}{r^2} \partial_\mu\gamma \left[\frac{1}{2}\partial_\mu\gamma (1-\mu^2)-\mu\right]\right\}\right], \\
 \frac{\tilde{S}_{\omega}}{r_{eq}r^2}(r,\mu)=&e^{(\gamma-2\tilde{\sigma})/2} \left[-16\pi e^{2\alpha}\frac{(\Omega-\omega)(\varepsilon+p)}{1-v^2}+\omega\left\{-8\pi e^{2\alpha}\frac{\left[(1+v^2)\varepsilon+2u^2p\right]}{1-v^2}
 -\frac{1}{r}\left(2\partial_r\tilde{\sigma}+\frac{1}{2}\partial_r\gamma \right)   \right.\right.     \nn\\
 &+\frac{1}{r^2}\mu\left(2\partial_\mu\tilde{\sigma}+\frac{1}{2}\partial_\mu\gamma\right)+\frac{1}{4}(4(\partial_r\tilde{\sigma})^2-(\partial_r\gamma)^2)+\frac{1}{4r^2}(1-\mu^2)(4(\partial_\mu\tilde{\sigma})^2-(\partial_\mu\gamma)^2)\nn\\
 &-\left.\left.r^2(1-\mu^2)e^{-2\tilde{\sigma}}\left[(\partial_r\omega)^2+\frac{1}{r^2}(1-\mu^2)(\partial_\mu\omega)^2\right]\right\}\right], \\
 \tilde{S}_{\phi}(s,\mu)=&-s^2(s-1)^2\p_s\gamma\p_s\phi-(1-\mu^2)\p_{\mu}\gamma\p_{\mu}\phi+4\pi k(\phi)\frac{r_{eq}^2 s^2}{(1-s)^2}e^{2\alpha}(\varepsilon-3p),
 \end{align}
 \end{widetext}
 where 
 \be v=(\Omega-\omega)r \sin\theta e^{-\tilde{\sigma}}, \ee
 is the proper velocity with respect to the zero angular momentum observers. 
 
 The calculated coefficients can then be used in the expressions given in Section \ref{sec:moments} to calculate the mass, angular momentum and scalar moments of the scalarized neutron star.

\bibliography{biblio}

\end{document}